\documentclass[preprint,showpacs,preprintnumbers,amsmath,amssymb]{revtex4-1}

\usepackage{graphicx}
\usepackage{dcolumn}
\usepackage{bm}

\begin{document}

\preprint{APS/XX0000}

\title{Neutrino-nucleus reactions on $^{16}$O based on new shell-model Hamiltonians}

\author{Toshio Suzuki$^{1,2}$, 
Satoshi Chiba$^{3}$, Takashi Yoshida$^{4}$, Koh Takahashi$^{5}$, 
Hideyuki Umeda$^{4}$  
}
\affiliation{$^{1}$ Department of Physics,
College of Humanities and Sciences, Nihon University     
Sakurajosui 3-25-40, Setagaya-ku, Tokyo 156-8550, Japan\\
$^{2}$ National Astronomical Observatory of Japan,
Mitaka, Tokyo 181-8588, Japan\\ 
\email{suzuki@chs.nihon-u.ac.jp}
%
$^{3}$ Research Laboratory for Nuclear Reactors, Tokyo Institute of Technology, Meguro, Tokyo 152-8550, Japan\\
$^{4}$ Department of Astronomy, Graduate School of Science, The University of Tokyo, Hongo, Bunkyo-ku, Tokyo 113-0033, Japan\\
$^{5}$ Argelander-Institute f\"{u}r Astronomie, Universit\"{a}t Bonn, D-53121 Bonn, Germany}

\date{\today}

\begin{abstract}
Neutrino-induced reactions on $^{16}$O are investigated by shell-model calculaions with new shell-model Hamiltonians, which can describe well the structure of $p$-shell and $p$-$sd$ shell nuclei. 
Distribution of the spin-dipole strengths in $^{16}$O, which give major contributions to the $\nu$-$^{16}$O reaction cross sections, is studied with the new Hamiltonians. Muon-capture reaction rates on $^{16}$O are also studied to discuss the quenching of the axial-vector coupling in nuclear medium.
Charged-current and neutral-current reaction cross sections are evaluated in various particle and $\gamma$ emission channels as well as the total ones at neutrino energies up to $E_{\nu}\approx$ 100 MeV.
Branching ratios for the various channels are obtained by the Hauser-Feshbach statistical model calculations, and 
partial cross sections for single- and multi-particle emission 
channels are evaluated.
The cross sections updated are compared with previous continuum random phase approximation (CRPA) calculations.
Effects of multi-particle emission channels on nucleosynthesis in core-collapse supernova (SN) explosions are investigated. 
Inclusion of $\alpha$p emission channels is found to lead to an enhancement of production yields of $^{11}$B and $^{11}$C through $^{16}$O ($\nu$, $\nu$' $\alpha$p) $^{11}$B 
and $^{16}$O ($\nu$, e$^{-}$ $\alpha$p) $^{11}$C reactions, respectively.
\end{abstract}

\maketitle


\def\be{\begin{equation}}
\def\ee{\end{equation}}
\def\bea{\begin{eqnarray}}
\def\eea{\end{eqnarray}}
\def\br{\bf r}


    



\section{INTRODUCTION}

Study of neutrino-nucleus reactions at neutrino energies up to $E_{\nu}$ =100 MeV is an important subject for the detection of supernova (SN) neutrinos.
Several nuclear targets such as $^{12}$C, $^{16}$O and $^{40}$Ar are especially of interest from the point of view of their availability.
Accurate evaluation of the $\nu$-induced cross sections is crucial for the study of neutrino production and nucleosythesis in SN explosions as well as neutrino oscillation properties.

Recently, new shell-model Hamiltonians become available due to the development of the study of exotic nuclei. Important roles of the tensor interaction are taken into account in the new Hamiltonians, so that they can explain shell evolutions and change of magic numbers toward driplines \cite{OSF,OSH}. 
Spin modes of nuclei such as Gamow-Teller (GT) transtions, magnetic dipole transitions and moments are found to be well described by these Hamiltonians.

Neutrino-induced reaction cross sections on $^{12}$C have been evaluated with a new $p$-shell Hamiltonian, SFO \cite{SFO}, which can reproduce the GT transition strength in $^{12}$C. 
The SFO is constructed to be used in $p$-$sd$ shell with the inclusion of excitations of $p$-shell nucleons to $sd$-shell up to 2-3 $\hbar\omega$ configurations. 
Exclusive and inclusive charged- and neutral-current reaction cross setions for DAR (decay-at-rest) $\nu$'s are found to be well reproduced by the SFO within experimental error bars \cite{SCY,SK}. 
Cross sections for particle emission channels obtained by Hauser-Feshbach method are used to study nucleosynthesis of light elements in SN explosions \cite{SCY,YSK}. Production yields of $^{7}$Li and $^{11}$B are found to be enhanced compared with previous calculations \cite{Woos}. 
In case of MSW (Mikheyev-Smirnov-Wolfenstein) resonance $\nu$ oscillations, the yield ratio of $^{7}$Li/$^{11}$B is pointed out to be a good measure of neutrino mass hierarchy \cite{YKM,YSK,SK}. 

The same Hamiltonian, SFO, is applied to $\nu$-induced reactions on $^{13}$C. 
The $^{13}$C isotope, whose abundance in carbon isotopes is 1.1$\%$, is a favorable target for detection of low energy $\nu$ below $E_{\nu}\sim$ 10 MeV, as reaction cross sections on $^{12}$C vanish below $E_{\nu}$ =15 MeV. 
The updated cross sections for $^{13}$C are found to be enhanced compared with those of Cohen-Kurath interactions \cite{SBK,FK}. 

Since liquid argon is an important target for the $\nu$ detection, accurate evaluation of $\nu$-induced cross sections on $^{40}$Ar is crucial for the study of SN neutrinos.
Charged-current cross sections on $^{40}$Ar have been evaluated with the use of recent new shell-model Hamiltonians.
Phenomenological Hamiltonians, SDPF-M \cite{Utsuno} and GXPF1J \cite{Honma}, are adopted for $sd$- and $pf$-shell, respectively, while the monopole-based universal interaction \cite{OSH} is used to obtain the $sd$-$pf$ cross shell matrix elements. 
The GT strength in $^{40}$Ar obtained in (p, n) reaction \cite{Bhat} is reproduced rather well by the shell-model calculations \cite{SH}. The GT strength is found to be large compared with a previous shell-model estimation \cite{Orm}.
The cross section for $^{40}$Ar is enhanced at $E_{\nu}<$ 50 MeV, where the GT transitions are dominant. 
Besides $1^{+}$ and $0^{+}$ transitions, cross sections are evaluated with random-phase approximation (RPA). The contributions from multipoles with 0$^{-}$, 1$^{-}$ and 2$^{-}$ become important at $E_{\nu}>$ 50 MeV. 
The cross section by the hybrid model is enhanced by about 20-40$\%$ compared to that obtained by RPA \cite{Kolbe} at lower energies, $E_{\nu}<$ 40 MeV. 

Here, we study $\nu$-induced reactions on $^{16}$O, which is the content of water target and is expected to be promising for SN neutrino detection \cite{KL}. 
In sect. 2, we investigate spin-dipole transition strengths in $^{16}$O based on new shell-model Hamiltonians, and discuss quenching of the axial-vector coupling of weak hadronic interaction in nuclear medium by studying muon-capture reaction on $^{16}$O.
In sect. 3, charged- and neutral-current reaction cross sections are evaluated by shell-model calculations. Total and partial cross sections for various particle and gamma emission channels are obtained and compared with previous calculations \cite{KL}.
In sect. 4, we discuss nucleosynthesis of $^{11}$B and $^{11}$C produced by emissions of $\alpha$ and proton from $^{16}$O in SN explosions. Summary is given in sect. 5.

\section{Spin-dipole strengths in $^{16}$O}

The shell-model Hamiltonian for $p$-shell, SFO, has been successfully applied to the GT transitions in $^{12}$C. The tensor components are properly improved in the $p$-shell part of SFO by enhancing the monopole terms in the spin-isospin flip channel.
In case of $^{16}$O, the GT transition gives only a minor contribution as $^{16}$O is an LS-closed shell nucleus. Dominant contributions come from spin-dipole (SD) transitions, where $p$-shell nucleons are excited into $sd$-shell.
Thus, the $p$-$sd$ cross shell interaction plays an important role.
The $p$-$sd$ shell cross shell part of SFO is a phenomenological Millener-Kurath interaction \cite{MK}. It is important to update the $p$-$sd$ cross shell part of the Hamiltonian with proper inclusion of the tensor interaction.
Here, we use a modified version of SFO, in which the tensor and two-body spin-orbit components of the $p$-$sd$ cross shell matrix elements are replaced by those of $\pi$+$\rho$ meson exchanges and $\sigma$+$\omega$+$\rho$ meson exchanges \cite{M3Y}, respectively. 
The modified Hamiltonian, referred as SFO-tls \cite{SO}, has been found to describe well the structure of neutron-rich carbon isotopes \cite{SO,RMP}. 

\begin{figure*}[tbh]
\hspace{-7mm}
\includegraphics[scale=1.02]{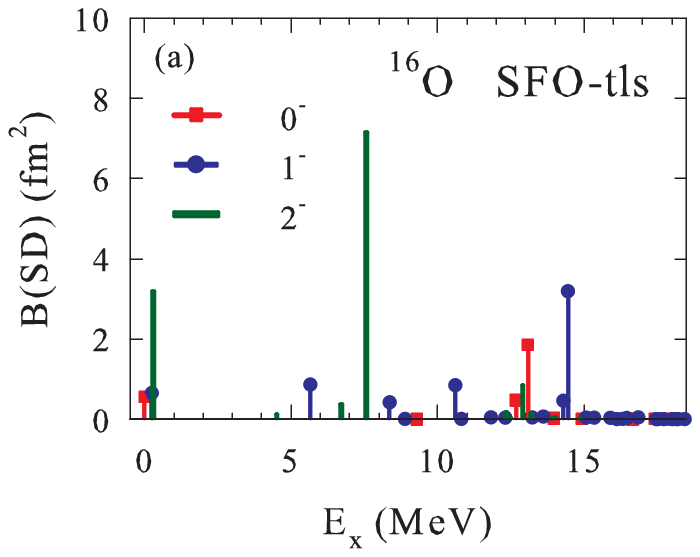}
\hspace{-7mm}
\includegraphics[scale=1.02]{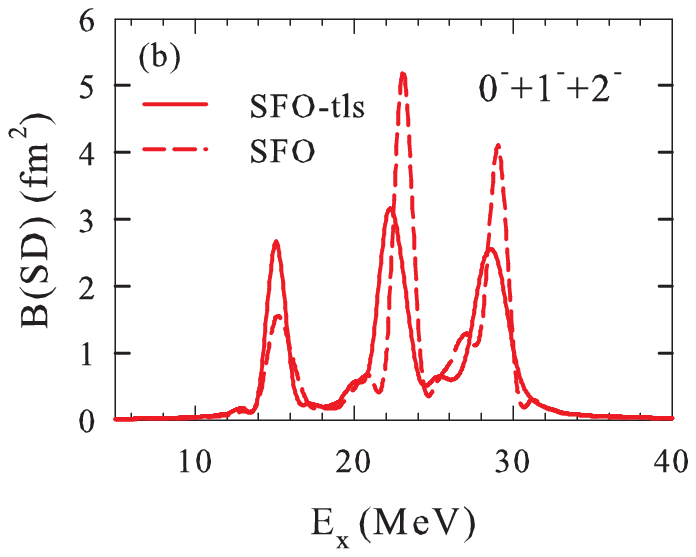}
\caption{(Color on line)
(a) Spin-dipole strengths for $^{16}$O $\rightarrow$ $^{16}$F obtained by shell-model calculations with SFO-tls. E$_x$ denotes the excitation energy of $^{16}$F.
(b) Comparison of the spin-dipole strengths for $^{16}$O $\rightarrow$ $^{16}$F obtained with SFO-tls and those with SFO. The strengths are folded over a Lorentzian with the width of 1 MeV. E$_x$ denotes the excitation energy of $^{16}$O.   
\label{fig:fig1}}
\end{figure*}

Spin-dipole strengths with $\lambda^{\pi}$ are defined as
\begin{eqnarray}
B(SD\lambda)_{\mp} &=& \frac{1}{2 J_i +1} \sum_{f} \mid<f\parallel S^{\lambda}_{\mp} \parallel i>\mid^{2} \nonumber\\
S^{\lambda}_{\mp,\mu} &=& r [Y^{1} \times \vec{\sigma}]^{\lambda}_{\mu} t_{\mp} 
\end{eqnarray}
where $J_i$ is the spin of the initial state and $t_{-}\mid \mbox{n}>$ =$\mid \mbox{p}>$ and $t_{+}\mid \mbox{p}>$ =$\mid \mbox{n}>$. 
Calculated spin-dipole strengths for $^{16}$O (0$_{g.s.}^{+}$) $\rightarrow$ $^{16}$F ($\lambda^{\pi}$ = 0$^{-}$, 1$^{-}$, 2$^{-}$) obtained with SFO-tls as well as SFO are shown in Fig. \ref{fig:fig1}.
Configuration up to 3$\hbar\omega$ (2$\hbar\omega$) excitations from $p$-shell to $sd$-shell are included for the spin-dipole (the ground) states. 
Experimental energies of the lowest 0$^{-}$, 1$^{-}$ and 2$^{-}$ states, $E_x$ = 0.0, 0.194 and 0.425 MeV, respectively \cite{NNDC}, are rather well reproduced by shell-model calculations with SFO-tls. 
Calculated energies are $E_x$ = 0.0, 0.251 and 0.299 MeV for 0$^{-}$, 1$^{-}$ and 2$^{-}$ states, respectively, which agree with the experimental values within 0.13 MeV.  
The 2$^{-}$ state at $E_x$ =7.50 MeV, which has the largest strength among the spin-dipole states, is located at the right position with $E_x$ =7.57 MeV for the SFO-tls. 
In case of the SFO, the calculated energy of the corresponding 2$^{-}$ state is $E_x$ =8.27 MeV, higher than the observed energy by 0.7 MeV. 
The energy of the lowest 2$^{-}$ state for the SFO is $E_x$ = 0.705 MeV, which is also higher than the experimental value by 0.28 MeV. 
The spin-dipole strengths for the SFO are shifted toward higher energy region by 0.3-1 MeV compared with those for the SFO-tls as shown in Fig. \ref{fig:fig1}(b). 

Sum rule values of the SD strength in $^{16}$O are given by using the harmonic oscillator wave functions to be \cite{SS},
\begin{equation}
S_{\lambda}(SD) = \sum_{\mu}\mid<\lambda, \mu\mid S^{\lambda}_{-,\mu} \mid 0>\mid^2
= \left\{ \begin{array}{rr}
\frac{3}{4\pi} 4\mbox{b}^2 =2.99 \mbox{fm}^2 &  \lambda^{\pi} =0^{-} \\
\frac{3}{4\pi} 12\mbox{b}^2 =8.98 \mbox{fm}^2 & \lambda^{\pi} =1^{-} \\
\frac{3}{4\pi} 20\mbox{b}^2 =14.96 \mbox{fm}^2 & \lambda^{\pi} =2^{-}  
\end{array} \right.
\end{equation}
where the excitations from $p$ to $sd$ shells are consideed and the oscilltor parameter is taken to be b =1.77 fm$^2$.
The SD strength for SFO-tls below $E_{x}$ =40 MeV exhausts 99$\%$, 77$\%$ and 82$\%$ of the sum rule values for 0$^{-}$, 1$^{-}$ and 2$^{-}$, respectively. 
As the tensor interaction shifts the 1$^{-}$ (0$^{-}$ and 2$^{-}$) strength to higher (lower) energy region, the summed fraction up to $E_{x}$ = 40 MeV for 1$^{-}$ is smaller than those for 0$^{-}$ and 2$^{-}$. 

The energy-weghted sum (EWS) of the spin-dipole operator Eq. (1) defined by
\begin{equation}
EWS^{\lambda}_{\pm} = \sum_{\mu} \mid<\lambda, \mu \mid S^{\lambda}_{\pm,\mu} \mid 0>\mid^2 (E_{\lambda}-E_{0}),
\end{equation}
is used to obtain
\begin{eqnarray}
EWS^{\lambda} &=& EWS^{\lambda}_{-} + EWS^{\lambda}_{+}\nonumber\\
&=& \frac{1}{2} <0 \mid [S^{\lambda}_{+}, [H, S^{\lambda}_{-}]] + [[S^{\lambda}_{-}, H], S^{\lambda}_{+} \mid 0>. 
\end{eqnarray}
The EWS rule value for the kinetic energy term ($K$) for $H$ = $\frac{p^2}{2m}$ with $m$ the nucleon mass is given as \cite{TS}
\begin{equation}
EWS^{\lambda}_{K} = \frac{3}{4\pi}(2\lambda+1)\frac{\hbar^2}{2m} A
[1 + \frac{f_{\lambda}}{3A} <0 \mid \sum_{i} \vec{\sigma}_{i}\cdot \vec{\ell}_{i} \mid 0>]
\end{equation}
where $f_{\lambda}$ = 2, 1 and -1 for $\lambda^{\pi}$ = 0$^{-}$, 1$^{-}$ and 2$^{-}$, respectively.  For LS-closed nuclei, the last term in Eq. (5) vanishes. 
The EWS rule value for the one-body spin-orbit potential ($LS$),
$V_{LS}$ = -$\xi \sum_{i}\vec{\ell}_{i}\cdot\vec{\sigma}_{i}$, is given as \cite{TS}
\begin{equation}
EWS^{\lambda}_{LS} = \frac{3}{4\pi}(2\lambda+1)\frac{f_{\lambda}}{3}\xi
<0 \mid \sum_{i} (r_i^2 + g_{\lambda}r_i^2 \vec{\ell}_i \cdot\vec{\sigma}_i) \mid 0>
\end{equation}
where $g_{\lambda}$ = 1 for $\lambda^{\pi}$ = 0$^{-}$, 1$^{-}$ and $g_{\lambda}$ = -7/5 for $\lambda^{\pi}$ = 2$^{-}$.  
Note that $EWS^{\lambda =2}_{LS}$ is reduced by the spin-orbit potential.  
For an LS-closed $^{16}$O, 
$<0 \mid \sum_{i} \vec{\ell}_i \cdot \vec{\sigma}_i \mid 0>$ =0, 
$<0 \mid \sum_{i} r_i^2 \vec{\ell}_i \cdot \vec{\sigma}_i \mid 0>$ =0 and $EWS^{\lambda}_{-}$ = $EWS^{\lambda}_{+}$. 
Then, with $\xi$ = 1.87 MeV \cite{BM}, $EWS^{\lambda}_{-}$ for $K+LS$ terms are obtained to be 56.4, 144.1 and 155.9 MeV$\cdot$fm$^2$ for 0$^{-}$, 1$^{-}$ and 2$^{-}$, resectively. 

Values of $EWS^{\lambda}_{-}$ obtained by the shell-model calculations including up to 3$\hbar\omega$ excitations are 73.0 (76.1), 173.2 (175.0) and 246.5 (258.2) MeV$\cdot$fm$^2$ for the SFO-tls (SFO) for $\lambda^{\pi}$ = 0$^{-}$, 1$^{-}$ and 2$^{-}$, respectively. The contributions up to $E_x$ $\approx$ 50 MeV are included. 
The shell-model values of $EWS^{\lambda}_{-}$ are enhanced compared with those of the $K+LS$ terms by 1.29 (1.35) for 0$^{-}$, by 1.20 (1.21) for 1$^{-}$ and by 1.58 (1.66) for 2$^{-}$ in case of the SFO-tls (SFO). 
These enhancements are caused by the contributions from two-body spin-dependent interactions, especially the tensor interaction in the shell model Hamiltoninans.
The enhancement factor is noticed to be large for 2$^{-}$, where the tensor interaction works attractive to shift the spin-dipole strength to lower energy region. 
This can be shown also from the centroid energy of the strength defined by 
$\bar{E_{\lambda}}$ = $EWS^{\lambda}_{-}$/NEWS$^{\lambda}_{-}$, where
$NEWS^{\lambda}_{-}$ is the calculated value of $S_{\lambda}$(SD).             
Values obtained for the SFO-tls (SFO) are $\bar{E_{0}}$ = 24.5 (25.8) MeV, 
$\bar{E_{1}}$ = 25.1 (25.2) MeV and $\bar{E_{2}}$ = 20.1 (21.0) MeV. 
Splitting of the centroid energies reflects attractve (repulsive) effects of the spin-dependent interaction in $\lambda^{\pi}$ = 0$^{-}$ and 2$^{-}$ ($\lambda^{\pi}$ = 1$^{-}$).  
The shift of $\bar{E_0}$ and $\bar{E_2}$ to lower energies by 0.7-0.9 MeV from SFO to SFO-tls comes from the effects of the tensor interaction properly taken into account in the $p$-$sd$ cross shell part in the SFO-tls.  

Next, we discuss quenching of the axial-vector coupling constant $g_A$ in nuclear medium in order to get a reliable evaluation of $\nu$-induced reaction cross sections on $^{16}$O. 
A quenching factor of f=$g_{A}^{eff}$/$g_A$ =0.95 close to 1.0 is obtained for the SFO from the study of the GT transition in $^{12}$C \cite{SFO}. 
Magnetic moments of $p$-shell nuclei are found to be systematically well reproduced with the use of this quenching factor for the isovector spin $g$-factor \cite{SFO}.
In case of the SFO-tls, almost the same quenching factor, f=0.96, is obtained to reproduce the GT strength in $^{12}$C. 
Here, we study muon-capture on $^{16}$O to obtain information on the quenching factor in the spin-dipole transitions.

The muon capture rate for $^{16}$O ($\mu$, $\nu_{\mu}$) $^{16}$N from the 1s Bohr atomic orbit is given as \cite{DW}
\begin{equation}
\omega_{\mu} = \frac{2G^2}{1+\nu/M_T} \mid\phi_{1s}\mid^2 \frac{1}{2J_i +1}
(\sum_{J=0}^{\infty} \mid<J_f \parallel M_J -L_J \parallel J_i>\mid^2 
+\mid<J_f \parallel T_{J}^{el} -T_{J}^{mag} \parallel J_i>\mid^2),
\end{equation}
where $G$ =$G_F$ cos $\theta_C$ is the weak coupling constant with $G_F$ the Fermi constant, $\theta_C$ is the Cabibbo angle, $\nu$ is the neutrino energy, $M_T$ is the target mass, and
\begin{equation}
\mid \phi_{1s}\mid^2 = \frac{R}{\pi} (\frac{m_{\mu}M_{T}}{m_{\mu}+M_{T}} Z\alpha)^3
\end{equation}
where $m_{\mu}$ the muon mass, $\alpha$ the fine-structure constant, $Z$ =8 and $R$ is a reduction factor to take into account the finite nuclear size effect.
A value of $R$ =0.79 is adopted for $^{16}$O \cite{DW}.
The transition matrix elements for the Coulomb ($M_J$), longitudinal ($L_J$), electric ($T_{J}^{el}$) and magnetic ($T_{J}^{mag}$) multipole operators with the multipolarities $J$ are evaluated at the neutrino energy $\nu$ for the weak hadronic currents.
In case of muon-capture reactions, there are sizable contributions from the pseudoscalar coupling term in the axial-vector current.  The pseudoscalar form factor obtained from pion-pole dominance of the induced pseudoscalar coupling and the Goldberger-Treiman relation is used; 
\begin{equation}
F_{P}(q_{\mu}^2) = \frac{2M_N}{q_{\mu}^2 +m_{\pi}^2} F_A(q_{\mu}^2) 
\end{equation}
where $M_N$ the nucleon mass, $m_{\pi}$ the pion mass, $q_{\mu}^2$ the four-momentum transfer, and $F_A$ is the axial-vector form factor of nucleon with $F_A$(0) = -1.26. Dominant contributions come from a region at $q_{\mu}^2$ = $m_{\mu}$(2$\nu$-$m_{\mu}$) $\approx$ (0.42 fm$^{-1}$)$^2$, for which g$_p$=$m_{\mu}F_P$ $\approx$ 7.5.    

Total muon capture rates for $^{16}$O obtained with f =$g_{A}^{eff}$/$g_A$ =0.95 are 10.21$\times$10$^{4}$ s$^{-1}$ and 11.20$\times$10$^{4}$ s$^{-1}$ for the SFO and SFO-tls, respectively. 
They agree with the experimental value, 10.26$\times$10$^{4}$ s$^{-1}$ \cite{SMR} within 10$\%$.
We thus find that nearly the same quenching factor f $\approx$0.95 as in the GT transitions can be used also for the spin-dipole transitions within 10$\%$ accuracy for the strength.
The quenching factor close to 1 has been also reported to be favorable for CRPA in Ref. \cite{KLV}.

\section{Neutrino-induced reaction cross sections on $^{16}$O}
\subsection{Total cross sections}

Neutrino-induced reaction cross sections are evaluated by using the multipole expansion of 
the weak hadronic currents, 
\begin{equation}
J^{C_{\mp}}_{\mu} = J^{V_{\mp}}_{\mu} +J^{A_{\mp}}_{\mu},
\end{equation}
for charged-current reactions ($\nu$, e$^{-}$) and ($\bar{\nu}$, e$^{+}$), and
\begin{equation}
J^{N}_{\mu} = J^{A_3}_{\mu} +J^{V_3}_{\mu} -2sin^{2}\theta_{W} J^{\gamma}_{\mu},
\end{equation}
for neutral-current reactions, ($\nu$, $\nu$') and ($\bar{\nu}$, $\bar{\nu}$'), where $J^{V}_{\mu}$ and $J^{A}_{\mu}$ are vector and axial-vector currents, respectively, and $J^{\gamma}_{\mu}$ is the electromagnetic vector current with $\theta_{W}$ being the Weinberg angle.  
The reaction cross sections are given as the sum of the matrix elements of the Coulomb, longitudinal, transverse electric and magnetic multipole operators for the vector and axial-vector currents \cite{DW,SCY}. 
Here, all the transition matrix elements with the multipolarities up to $\lambda$=4 are taken into account with the use of harmonic oscillator wave functions.
Initial and final states are obtained by the shell-model calculations in the $p$-$sd$ shell configurations including up to 2 (3) $\hbar\omega$ excitations for the positive (negative) parity transitions. Excited states up to $E_{x}\approx$ 50 MeV are taken into account. 

\begin{figure*}[tbh]
\begin{center}
\hspace{-7mm}
\includegraphics[scale=1.02]{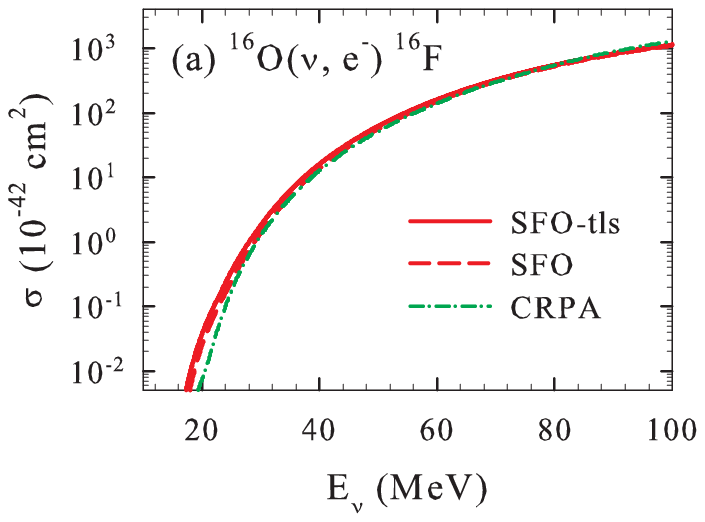}
\hspace{-8mm}
\includegraphics[scale=1.02]{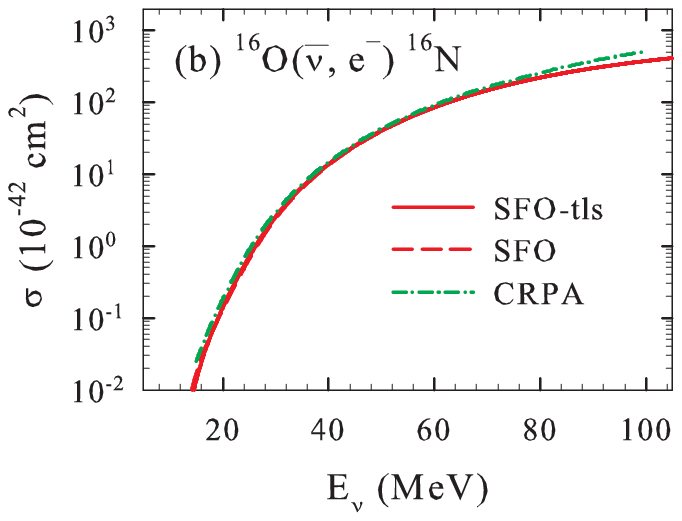}
\vspace{-7mm}
\includegraphics[scale=1.02]{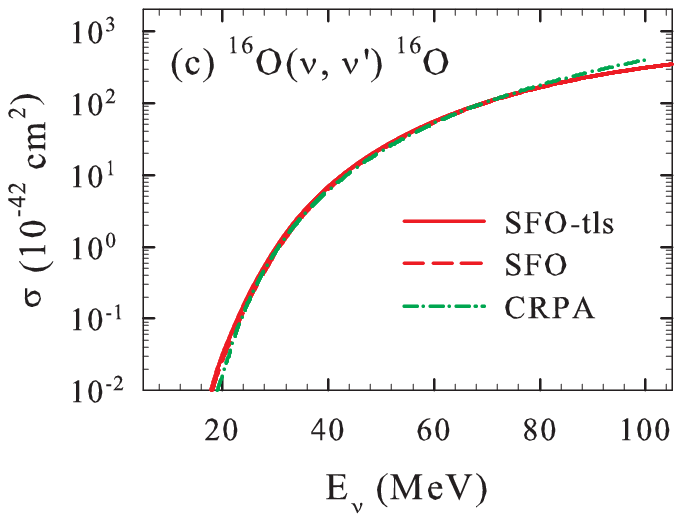}
\caption{(Color on line)
Total reaction cross sections for (a) $^{16}$O ($\nu$, e$^{-}$) $^{16}$F, (b) $^{16}$O ($\bar{\nu}$, e$^{+}$) $^{16}$N and (c) $^{16}$O ($\nu$, $\nu$') $^{16}$O obtained by shell-model calculations with the SFO-tls and SFO as well as those of the CRPA calculation \cite{KL}.
\label{fig:fig2}}
\end{center}
\end{figure*}

First, we show results of total cross sections for charged- and neutral-current reactions; $^{16}$O ($\nu_e$, e$^{-}$) $^{16}$F, $^{16}$O ($\bar{\nu}_e$, e$^{+}$) $^{16}$N and $^{16}$O ($\nu$, $\nu$') $^{16}$O. 
Calculated cross sections for $^{16}$O ($\nu_e$, e$^{-}$) $^{16}$F obtained with SFO-tls, SFO and CRPA (continuum-RPA)\cite{KL} are shown in Fig. \ref{fig:fig2}(a) as function of neutrino energy $E_{\nu}$. 
The quenching factor for the axial-vector coupling constant is taken to be $g_A^{eff}$/$g_A$ =0.95 for SFO-tls and SFO \cite{SFO}. 
 Calculated cross sections are summarized in Table \ref{tab:table1}. 
The cross section for the SFO-tls is found to be enhanced compared with the CRPA at $E_{\nu}<$ 80 MeV, and more than 50$\%$ at $E_{\nu}\leq$ 30 MeV, while it becomes smaller than the CRPA at $E_{\nu}>$ 90 MeV.
The cross section for the SFO is close to the CRPA at $E_{\nu}$ =30-80 MeV.
The enhancement of the cross section for the SFO-tls can be attributed to the shift of the SD strength to lower excitation energy region compared to the SFO and CRPA.
The cross sections for the SFO-tls and SFO are reduced compared to the CRPA at high $E_{\nu}$ as the CRPA calculations can take into account excited states wth higher excitation energies than the shell-model calculations of the present $p$-$sd$ shell configurations.   
 The cross sections averaged over neutrino spectra of Fermi distributions with temperature $T$ with zero-chemical potential are given in Table \ref{tab:table2}. 
The ratio of the cross sections $\sigma$(SFO-tls)/$\sigma$(CRPA) are 1.36 (1.12) for $T$ = 4 (8) MeV for $^{16}$O ($\nu_e$, e$^{-}$) $^{16}$F.

Total cross sections for $^{16}$O ($\bar{\nu}_e$, e$^{+}$) $^{16}$N and $^{16}$O ($\nu$, $\nu$') $^{16}$O obtained with the SFO-tls, SFO as well as CRPA are shown in Fig. \ref{fig:fig2}.
Averaged values of ($\nu$, $\nu$') and ($\bar{\nu}$, $\bar{\nu}$') cross sections are shown for the neutral-current reaction. The elastic coherent scattering is not included. Numerical values are given in Table \ref{tab:table1}. 
The cross sections folded over the neutrino spectra of Fermi distributions are given in Table \ref{tab:table2}. 
The charged-current ($\bar{\nu}$, e$^{+}$) cross sections for SFO-tls are close to those of SFO, and a bit smaller than those of CRPA by 10-20$\%$ at $E_{\nu}$ = 25-90 MeV.
The neutral-current reaction cross sections for SFO-tls and CRPA are close to each other and differ only by less than 10$\%$ at $E_{\nu}$ =30-80 MeV.

We next investigate cross sections for separate multipolarities as a function of excitation energies $E_{x}$ of the final states.
Cross sections for transitions with $\lambda^{\pi}$ =0$^{-}$, 1$^{-}$ and 2$^{-}$ as well as total ones for ($\nu_e$, e$^{-}$), ($\bar{\nu}_e$, e$^{+}$) and ($\nu$, $\nu$') reactions obtained with SFO-tls are shown in Figs. 3--5 for $E_{\nu}$ = 30 MeV and 50 MeV.

\begin{figure*}[tbh]
\begin{center}
\hspace{-7mm}
\includegraphics[scale=1.03]{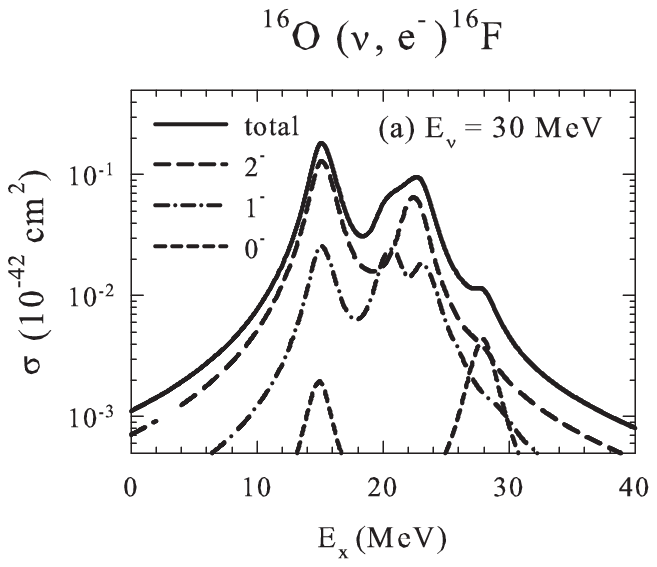}
\hspace{-10mm}
\includegraphics[scale=1.03]{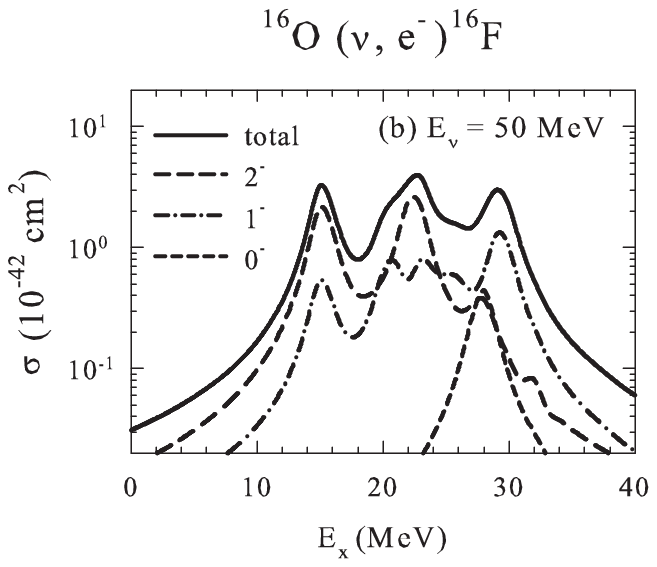}
\vspace{-5mm}
\caption{
Contributions from each 0$^{-}$. 1$^{-}$ and 2$^{-}$ multipoles as well as the total cross sections for $^{16}$ ($\nu$, e$^{-}$) $^{16}$F as function of excitation energy $E_{x}$ at (a) $E_{\nu}$ = 30 MeV and (b) $E_{\nu}$ = 50 MeV.
The cross sections are folded over a Lorentian with the width of 1 MeV.
\label{fig:fig3}}
\end{center}
\end{figure*}

\begin{figure*}[tbh]
\begin{center}
\hspace{-7mm}
\includegraphics[scale=1.03]{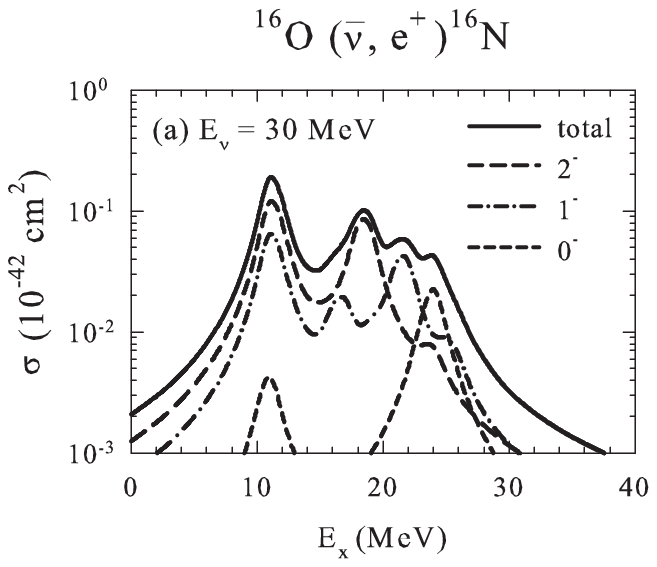}
\hspace{-10mm}
\includegraphics[scale=1.03]{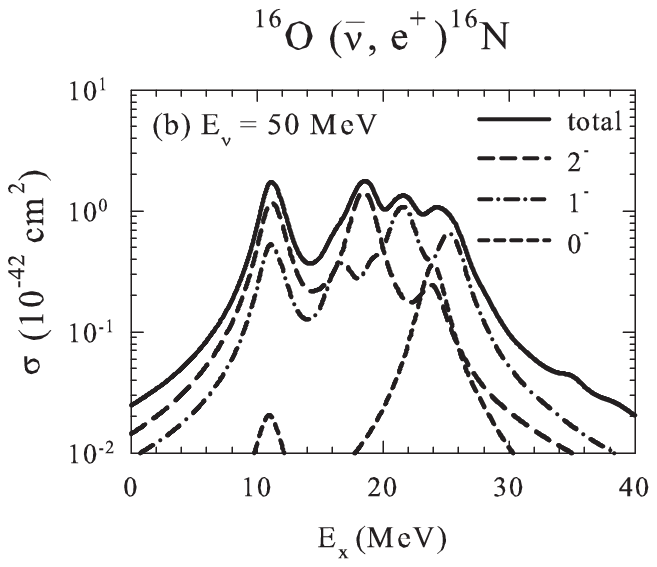}
\vspace{-5mm}
\caption{
The same as Fig. 3 for $^{16}$O ($\bar{\nu}$, e$^{+}$) $^{16}$N. 
\label{fig:fig4}}
\end{center}
\end{figure*}

\begin{figure*}[tbh]
\begin{center}
\hspace{-7mm}
\includegraphics[scale=1.03]{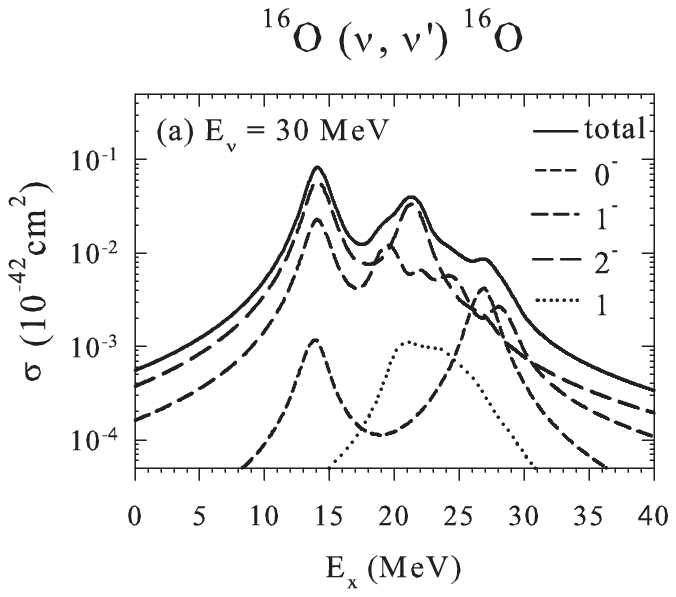}
\hspace{-10mm}
\includegraphics[scale=1.03]{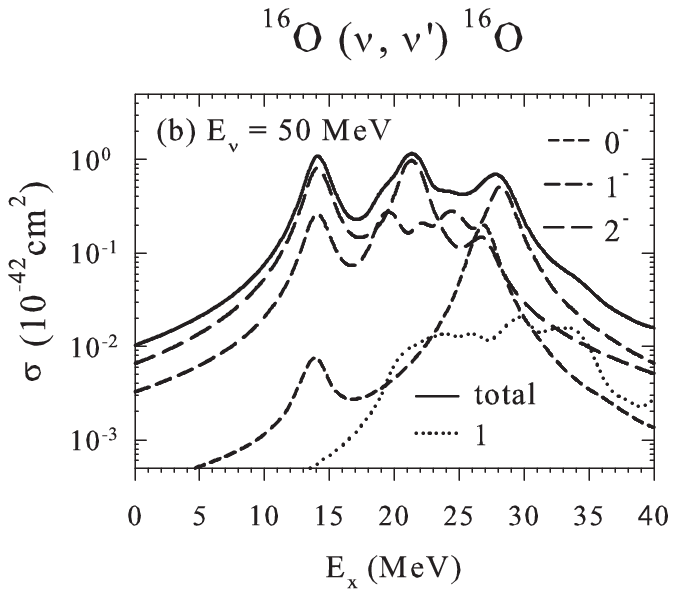}
\vspace{-5mm}
\caption{
The same as Fig. 3 for $^{16}$O ($\nu$, $\nu$') $^{16}$O.
Contributions from the 1$^{+}$ multipole are also shown. 
\label{fig:fig5}}
\end{center}
\end{figure*}

 We see from Figs. 3--5 that dominant contributions come from 2$^{-}$ transitions at $E_{x}\lesssim$ 25 MeV.
The contributions from 0$^{-}$ transitions become important around $E_{x}\sim$ 27 MeV for $E_{\nu}$ = 30 MeV, but the contributions from 1$^{-}$ become dominant around $E_{x}\sim$ 28 MeV for $E_{\nu}$ =50 MeV.
These behaviors reflect the SD strengths shown in Fig. \ref{fig:fig1}(a); large low-lying 2$^{-}$ strength at $E_{x}$ = 0 and 7.5 MeV, 0$^{-}$ strength at $E_{x}\sim$ 13 MeV and 1$^{-}$ strength at $E_{x}\sim$ 14.4 MeV.
Contributions from the GT (1$^{+}$) transitions are rather minor as shwon in Fig. \ref{fig:fig5}. They can, however, become comparable to the contributions from 0$^{-}$ (and 1$^{-}$) transitions at certain $E_{x}$ for $E_{\nu}$ = 30 (50) MeV.  
      
\subsection{Cross sections for particle and $\gamma$ emission channels}

Partial cross sections for various particle and $\gamma$ emission channels are evaluated by calculating the branching ratios from each excited level by the Hauser-Feshbach statistical model \cite{HF}. 
Single- and multi-particle decay channels involving neutron, proton, deutron, $\alpha$, $^{3}$He, $^{3}$H and $\gamma$ are considered.  All the levels obtained in the present shell-model calculations are adopted as levels in the decaying and daughter nuclei with specific isospin assignments. 
The particle transmission coefficients are calculated by the optical model \cite{Walt,Avr}. 
Isospin is conserved in the calculations, and possible isospin mixings are neglected.
The $\gamma$-transmission coefficients are calculated with the Brink's formula.
The E1 (electric dipole) and M1 (magnetic dipole) parameters are taken from the RIPL-2 database \cite{RIPL}. 
The $\gamma$ cascade in the initial excited nuclei and subsequent decays are fully considered.

\begin{figure*}[tbh]
\begin{center}
\hspace{-7mm}
\includegraphics[scale=1.03]{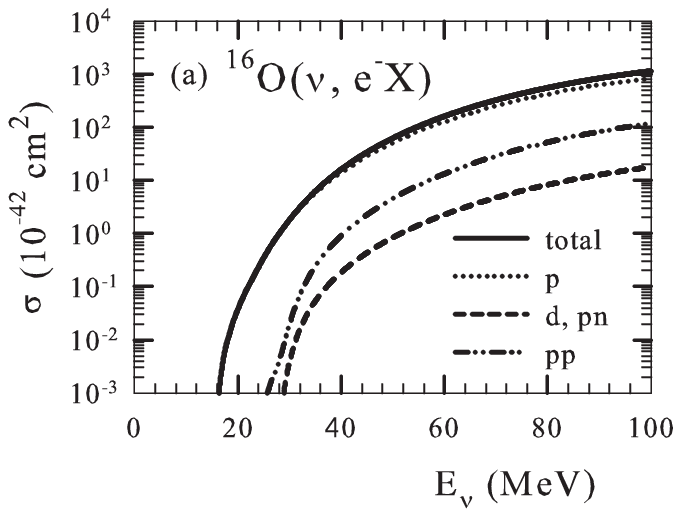}
\hspace{-10mm}
\includegraphics[scale=1.03]{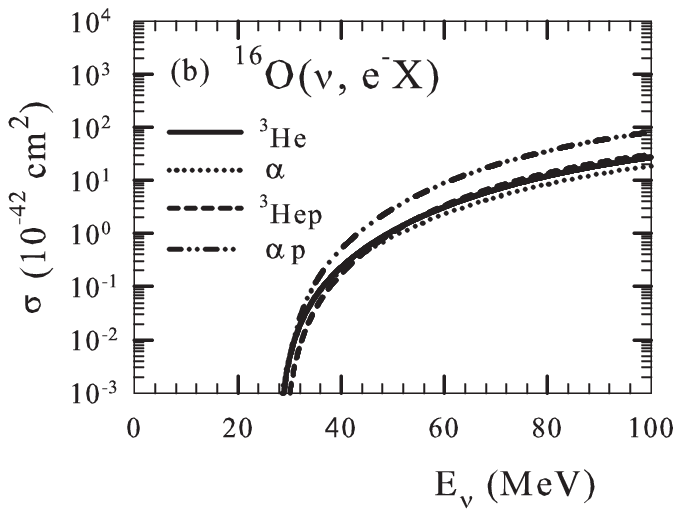}
\vspace{-5mm}
\caption{
Partial cross sections to various channels of $^{16}$O ($\nu$, e$^{-}$ X) as function of neutrino energy $E_{\nu}$.
(a) Cases for X =p, d and pn, and pp as well as the total cross section are shown.
(b) Cases for X =$^{3}$He (and dp, ppn), $\alpha$, $^{3}$He p, and $\alpha$ p are shown.
\label{fig:fig6}}
\end{center}
\end{figure*}

\begin{figure*}[tbh]
\begin{center}
\hspace{-7mm}
\includegraphics[scale=1.03]{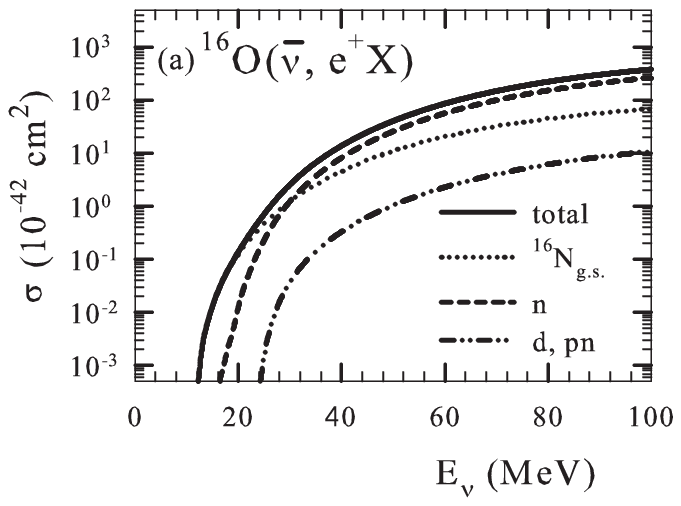}
\hspace{-10mm}
\includegraphics[scale=1.03]{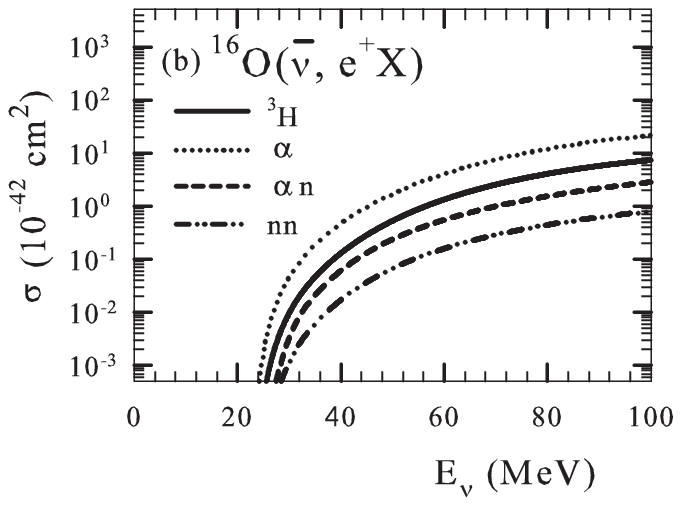}
\vspace{-5mm}
\caption{
The same as Fig. 6 for $^{16}$O ($\bar{\nu}$, e$^{+}$X).
(a) Cases for X =n, d and pn, and the transition to $^{16}$N$_{g.s.}$ as well as the total cross section are shown
(b) Cases for X =$^{3}$H (and dn, pnn), $\alpha$, $\alpha$ n, and nn are shown.
\label{fig:fig7}}
\end{center}
\end{figure*}

\begin{figure*}[tbh]
\begin{center}
\hspace{-7mm}
\includegraphics[scale=1.03]{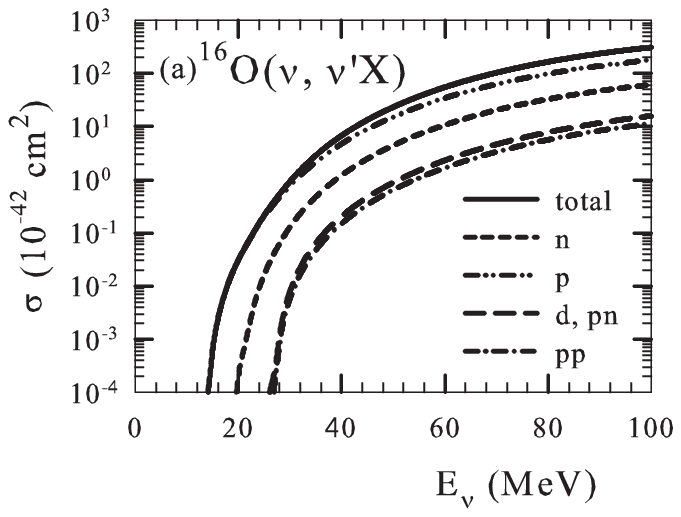}
\hspace{-10mm}
\includegraphics[scale=1.03]{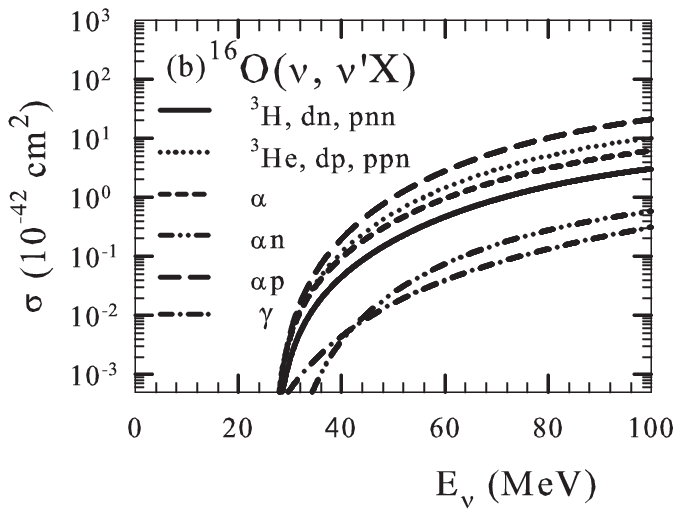}
\vspace{-5mm}
\caption{
The same as Fig. 6 for $^{16}$O ($\nu$, $\nu$'X).
(a) Cases for X =n. p. d and pn, and pp as well as the total cross section are shown.
(b) Cases for X =$^{3}$H (and dn, pnn), $^{3}$He (and dp, ppn), $\alpha$, $\alpha$ n, $\alpha$ p, and $\gamma$ are shown. 
\label{fig:fig8}}
\end{center}
\end{figure*}

\begin{figure*}[tbh]
\begin{center}
\hspace{-7mm}
\includegraphics[scale=1.02]{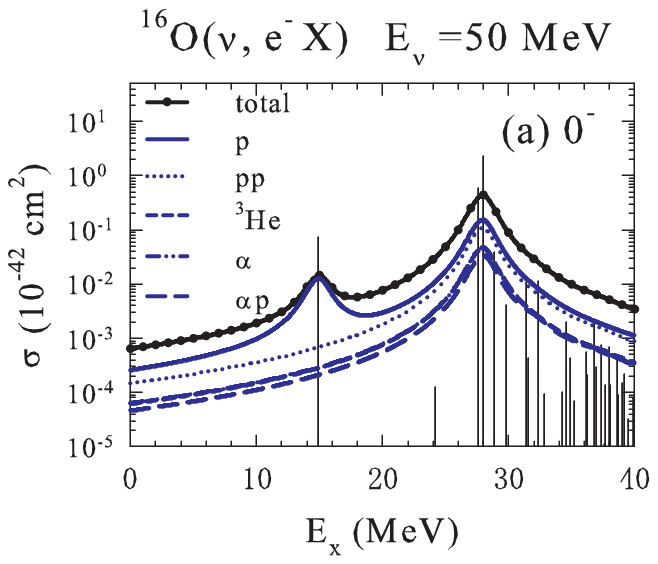}
\hspace{-10mm}
\includegraphics[scale=1.02]{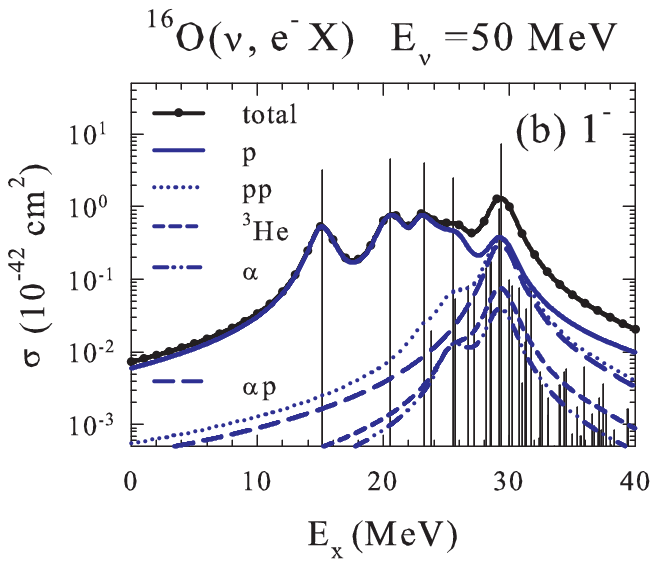}
\hspace{-10mm}
\includegraphics[scale=1.02]{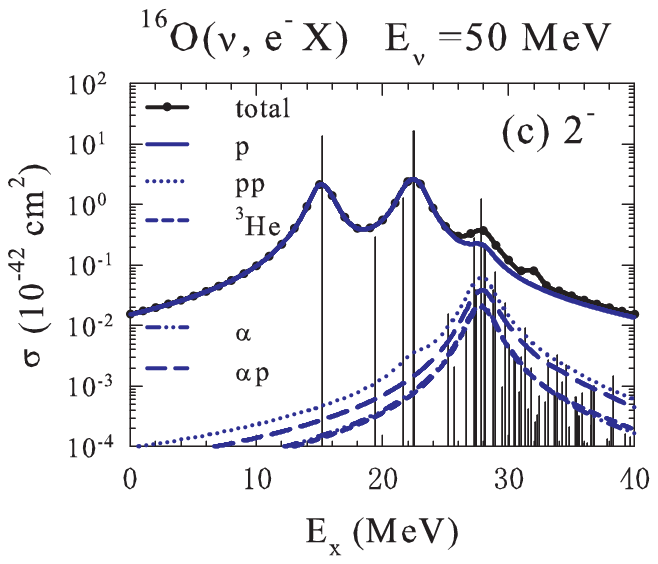}
\vspace{-10mm}
\caption{(Color on line)
Partial cross sections for $^{16}$O ($\nu$, e$^{-}$X) for each multipole (a) 0$^{-}$, (b) 1$^{-}$ and 2$^{-}$ as function of $E_{x}$ for $E_{\nu}$ = 50 MeV.
Cases for X =p, pp, $^{3}$He, $\alpha$, and $\alpha$ p as well as the total cross section are shown.
\label{fig:fig9}}
\end{center}
\end{figure*}

\begin{figure*}[tbh]
\begin{center}
\hspace{-7mm}
\includegraphics[scale=1.02]{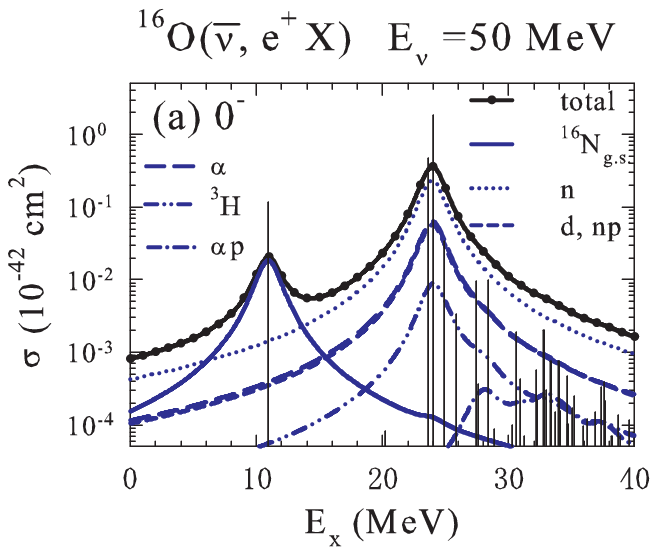}
\hspace{-10mm}
\includegraphics[scale=1.02]{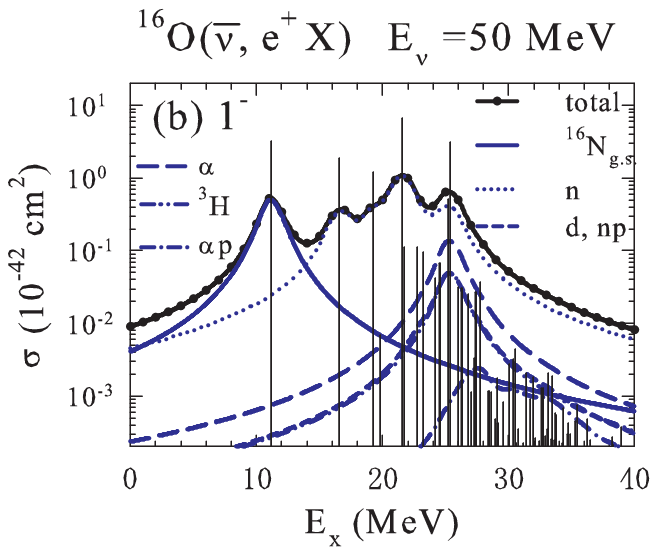}
\hspace{-10mm}
\includegraphics[scale=1.02]{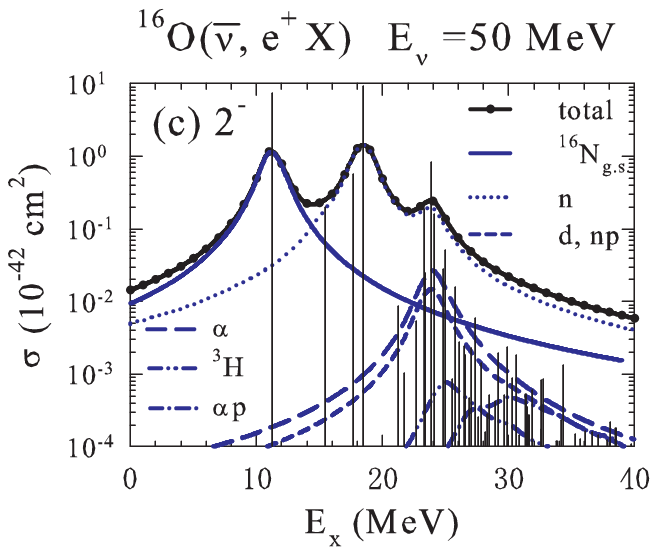}
\vspace{-10mm}
\caption{(Color on line)
The same as Fig. 9 for $^{16}$O ($\bar{\nu}$, e$^{+}$X). 
Cases for X = n, d or pn, $\alpha$, $^{3}$H, $\alpha$ p, and the transition to $^{16}$O$_{g.s.}$ are shown.  
\label{fig:fig10}}
\end{center}
\end{figure*}

\begin{figure*}[tbh]
\begin{center}
\hspace{-7mm}
\includegraphics[scale=1.02]{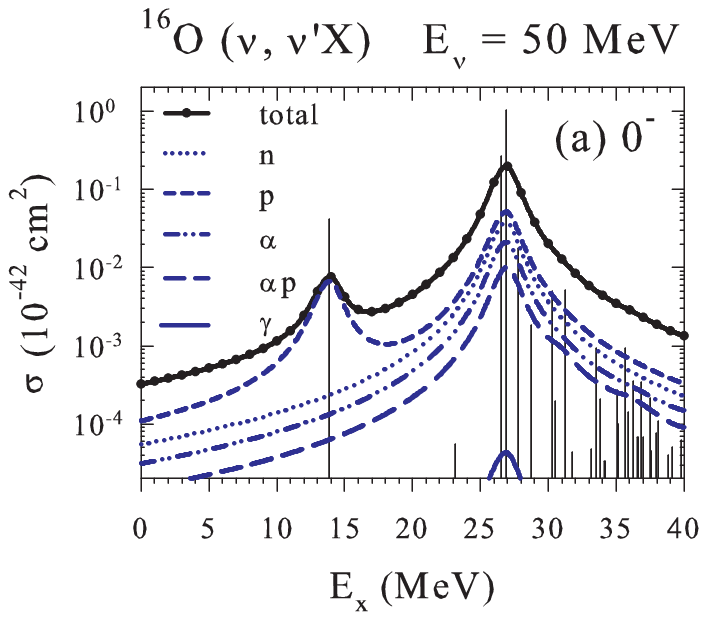}
\hspace{-8mm}
\includegraphics[scale=1.02]{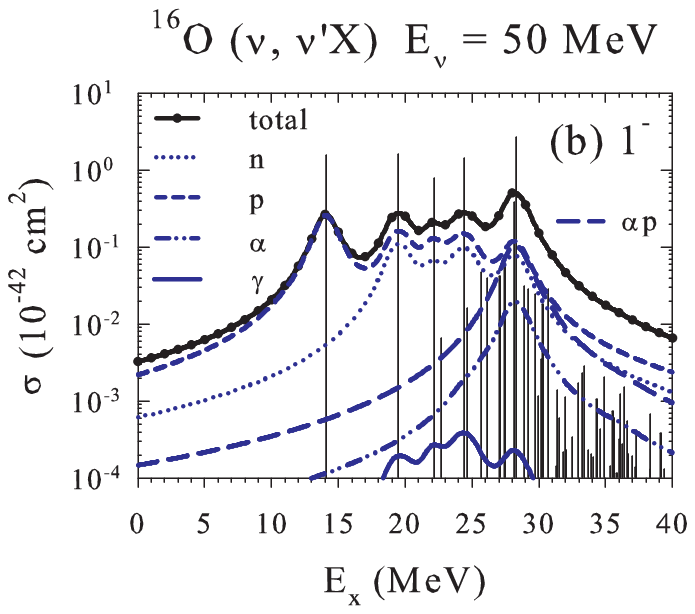}
\hspace{-9mm}
\includegraphics[scale=1.02]{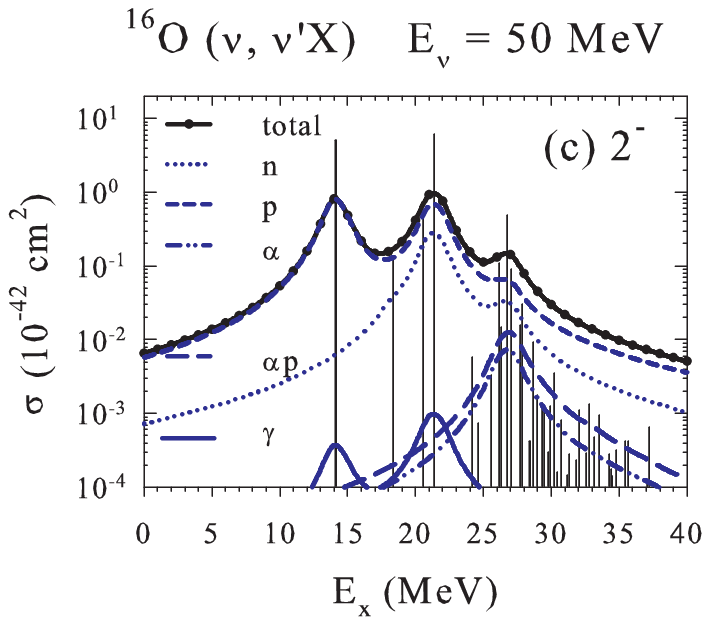}
\vspace{-5mm}
\caption{(Color on line)
The same as Fig. 10 for $^{16}$O ($\nu$, $\nu$'X).
Cases for X =n, p, $\alpha$, $\alpha$ p, and $\gamma$ are shown.
\label{fig:fig11}}
\end{center}
\end{figure*}

Calculated partial cross sections for various channels obtained with the SFO-tls for $^{16}$O ($\nu_e$, e$^{-}$X), $^{16}$O ($\bar{\nu}_e$, e$^{+}$X) and $^{16}$O ($\nu$, $\nu$'X) reactions are shown in Figs. \ref{fig:fig6}, \ref{fig:fig7} and \ref{fig:fig8}, respectively.
For $^{16}$O ($\nu_e$, e$^{-}$X) and $^{16}$O ($\nu$, $\nu$'X) reactions, the proton emission channel gives the dominant contribution while for $^{16}$O ($\bar{\nu}_e$, e$^{+}$X) reaction the neutron emission channel and the transition to the ground state of $^{16}$N give the dominant contributions.
Cross sections for various channels are given for SFO-tls in Tables III--V. 
Averaged cross sections folded over neutrino spectra of Fermi distributions are given in Tables VI--VIII, and compared with those of CRPA.
The cross sections of single proton emission channel in ($\nu$, e$^{-}$) reaction are enhanced by about 30$\%$-40$\%$  compared with the CRPA, while 
the cross sections of single neutron emission channel in ($\bar{\nu}$, e$^{+}$) reaction are reduced by about 20$\%$ compared with the CRPA.
In case of ($\nu$, $\nu$') reaction, the cross sections for the single proton (neutron) emission are enhanced (reduced) by 10$\%$ (25$\%$) compared with th CRPA, but the sum of the cross sections of the proton and neutron emissions are close to those of the CRPA within 2$\%$.
The cross sections of the $\alpha$ emission channel are enhanced by about 1.7-2.2 and 2.0-2.8 times compared with the CRPA in ($\nu$, e$^{-}$) and ($\bar{\nu}$, e$^{+}$) reactions, respectively.    
The contributions from the $\alpha$p emission channel are comparable to those from pp and pn emission channels in ($\nu$, e$^{-}$X) and ($\nu$, $\nu$'X) reactions, and large compared with those of the CRPA calculation \cite{KL}.

Partial cross sections for separate multipoles with 0$^{-}$, 1$^{-}$ and 2$^{-}$ are also shown in Figs. 9--11 as function of excitation energy $E_{x}$ for $E_{\nu}$ =50 MeV.    
Contributions other than single proton or neutron emissions become important at higher excitation energy region, $E_{x}$ = 25-30 MeV.   
Relatively large contributions from pp, $\alpha$p and $\alpha$ emission channels  are noticed around $E_{x}$ =30 MeV in the ($\nu$, e$^{-}$) reaction.
In the ($\bar{\nu}$, e$^{+}$) reaction, the contributions from $\alpha$ and d (pn) emission channels become important around $E_{x}$ =25 MeV.
Large contributions from the $\alpha$p emission channel are also noticed at $E_{x}$ =25-30 MeV in the ($\nu$, $\nu$') reaction.

\section{Nucleosynthesis of $^{11}$B and $^{11}$C in SN explosions}

In sect. 3, partial cross sections in various single- and multi-particle emission channels in $\nu$-$^{16}$O reactions are evaluated with the SFO-tls.
Here, we study effects of the multi-particle emission channels in nucleosynthesis of light elements in core-collapse SN explosions. 
Comparing the cross section for $^{16}$O ($\nu$, $\nu'\alpha$p) $^{11}$B with that for $^{12}$C ($\nu$, $\nu$'p) $^{11}$B which is the dominant reaction producing $^{11}$B in SN explosions \cite{SCY,YSK}, its ratio becomes as large as about 10$\%$ (compare Table \ref{tab:table8} and Fig. 2 of Ref. \cite{YSK} at $T$ = 4 and 8 MeV).
Thus, the reaction $^{16}$O ($\nu$, $\nu'\alpha$p) $^{11}$B is not negligible for nucleosynthesis of $^{11}$B in SN explosions.

\begin{figure*}[bht]
\includegraphics[scale=0.30,angle=-90]{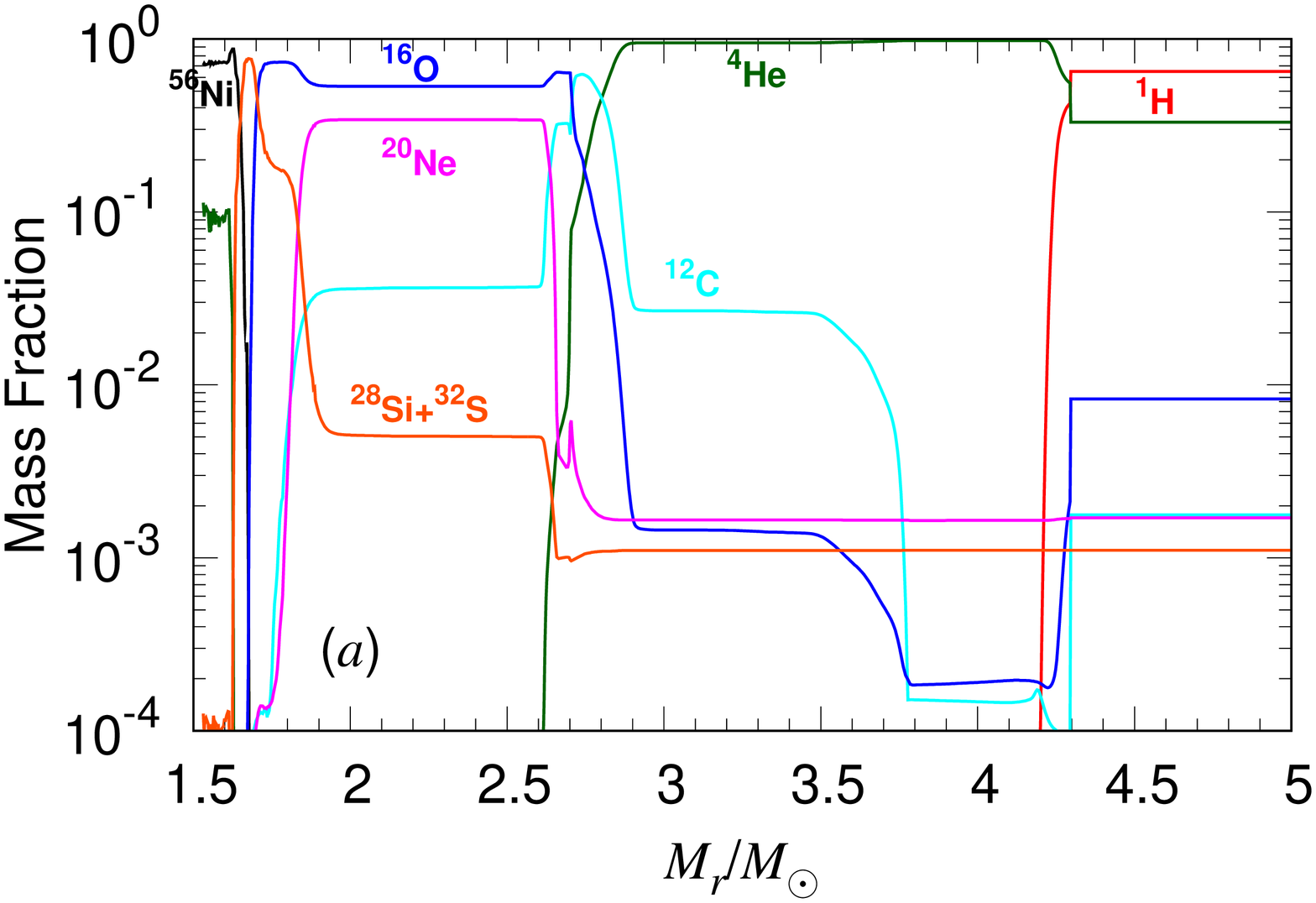}
\includegraphics[scale=0.30,angle=-90]{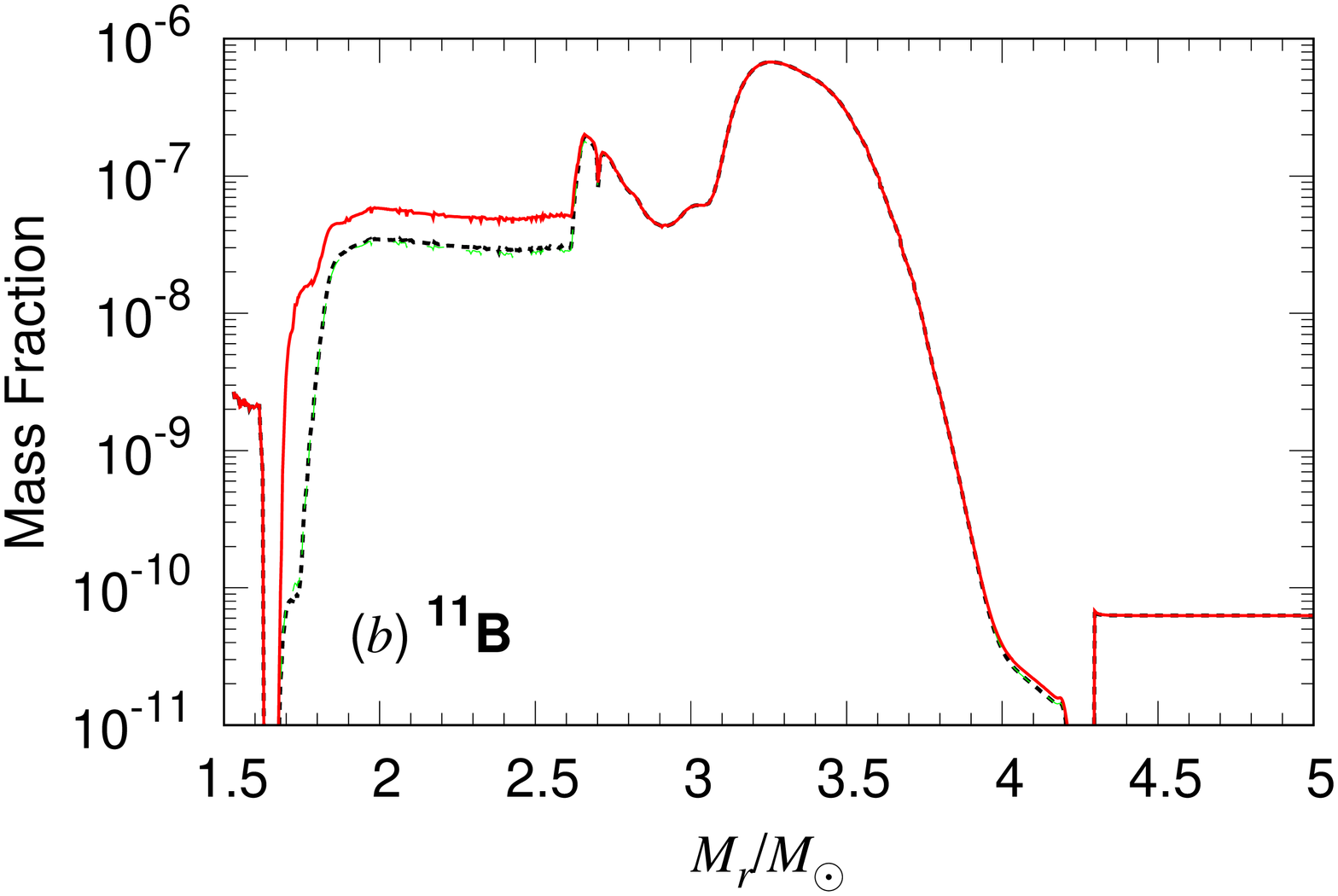}
\includegraphics[scale=0.30,angle=-90]{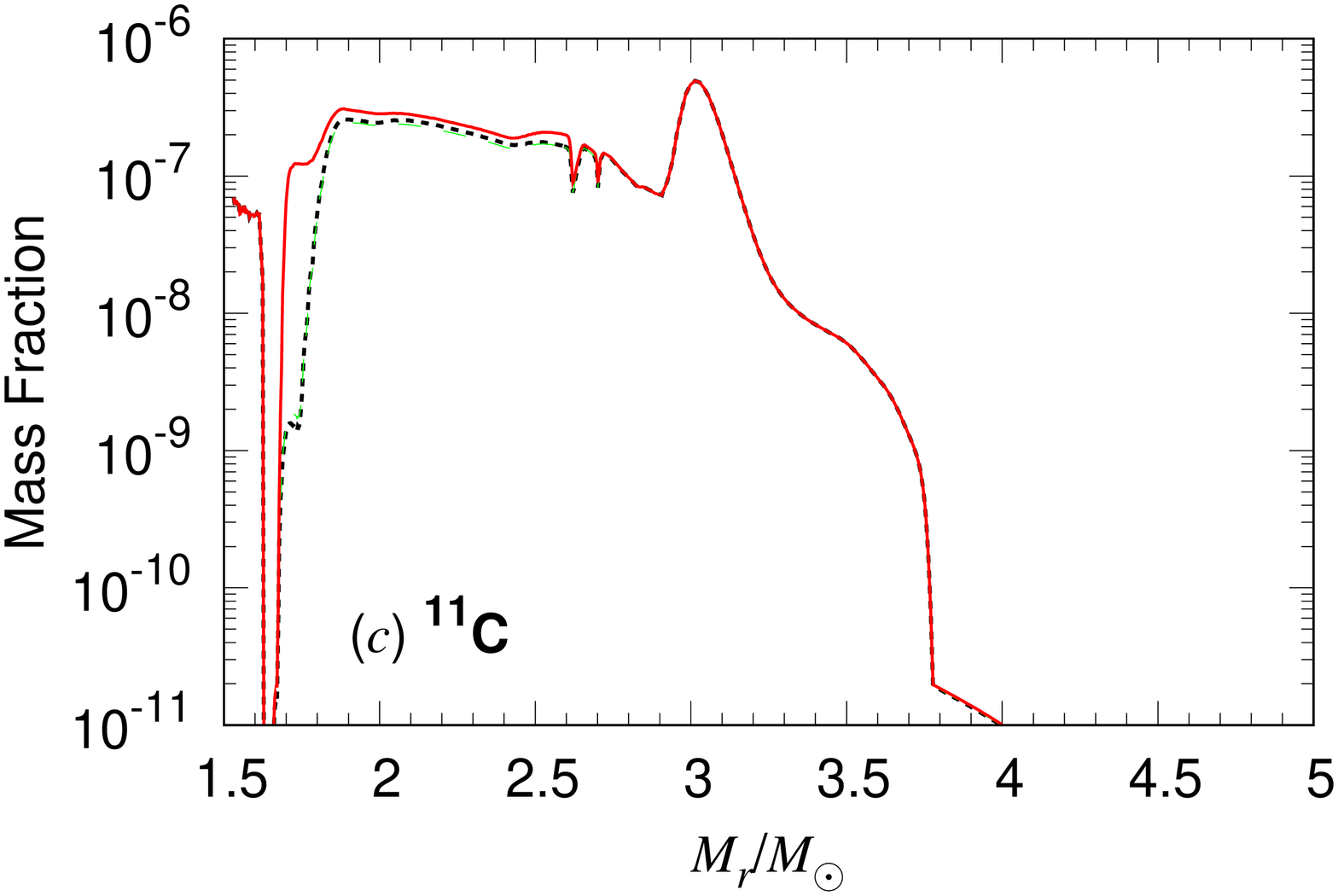}
\includegraphics[scale=0.30,angle=-90]{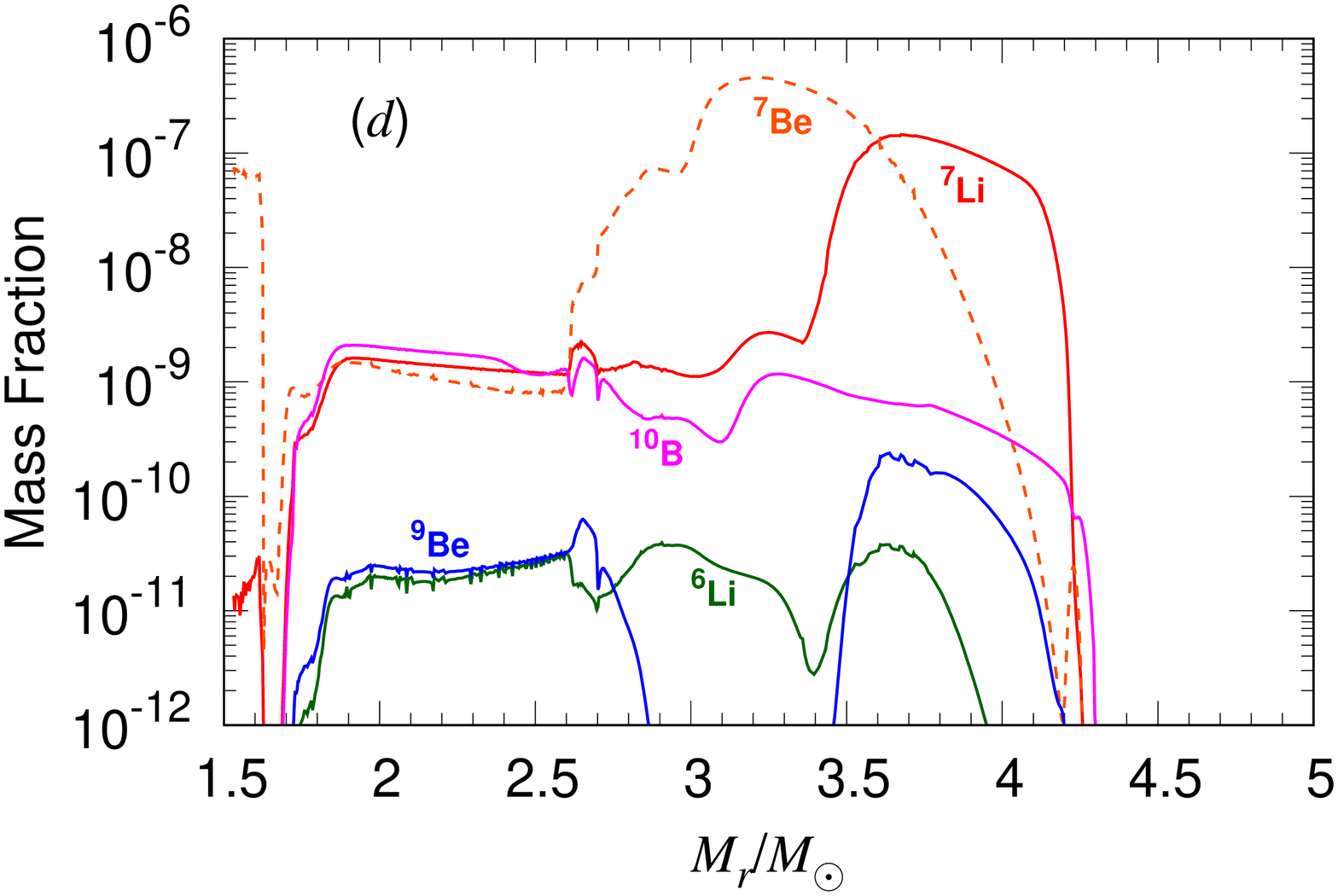}
\caption{(Color on line)  Mass fraction distributions of (a) main elements, (b) $^{11}$B, (c) $^{11}$C and (d) $^{6}$Li, $^{7}$Li, $^{7}$Be $^{9}$Be and $^{10}$B for a model with $M$ = 15$M_{\odot}$. 
The horizontal axis denotes the cumulative mass of the star from the center in the solar mass unit $M_{\odot}$. 
Shot-dashed, long-dashed and solid curves in (b) and (c) are for the case 1, case 2, and case 3, respectively (see text). Curves in (d) are for the case 3.
\label{fig:fig12}}
\end{figure*} 

\begin{figure}[bht]
\includegraphics[scale=0.30,angle=-90]{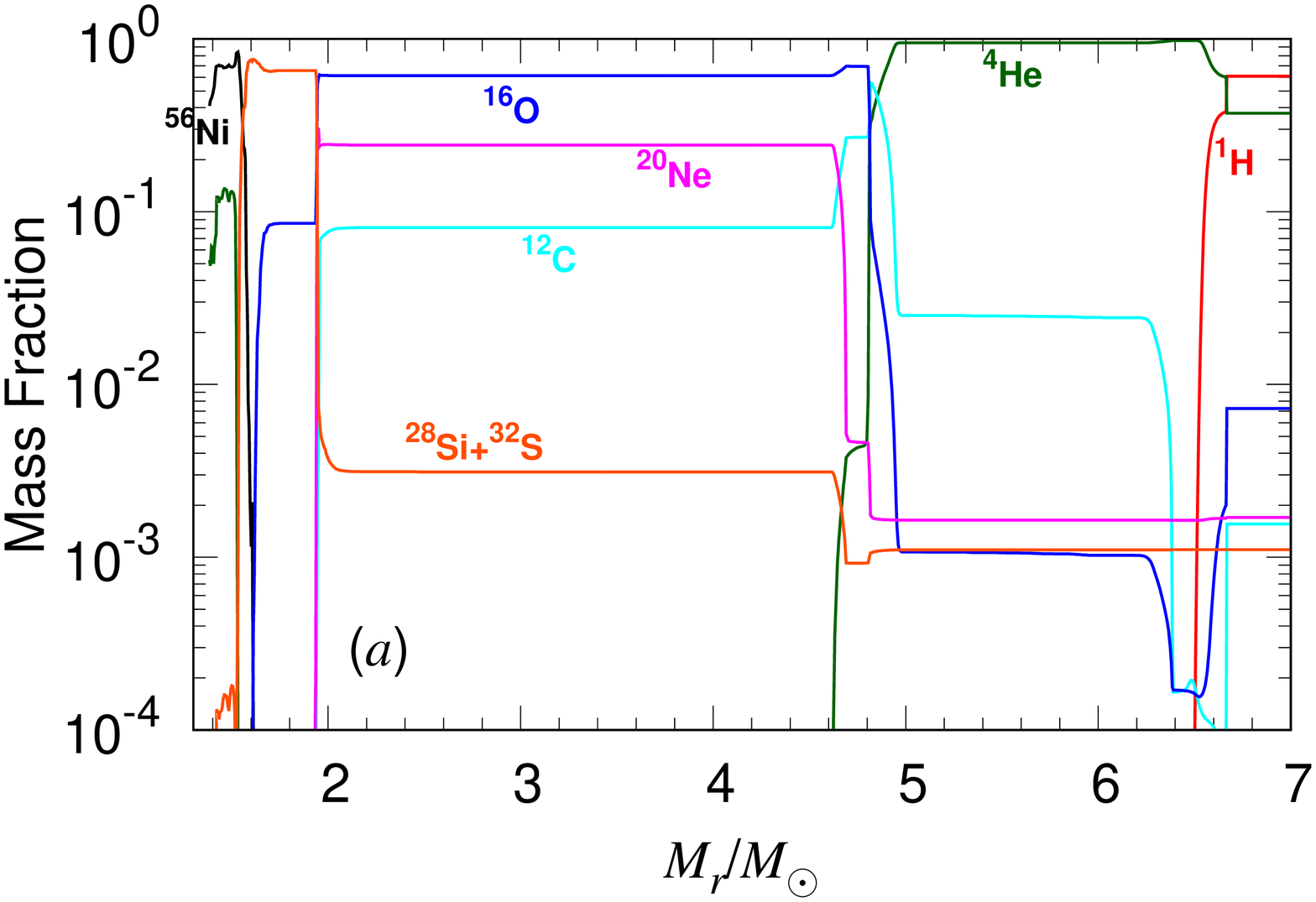}
\includegraphics[scale=0.30,angle=-90]{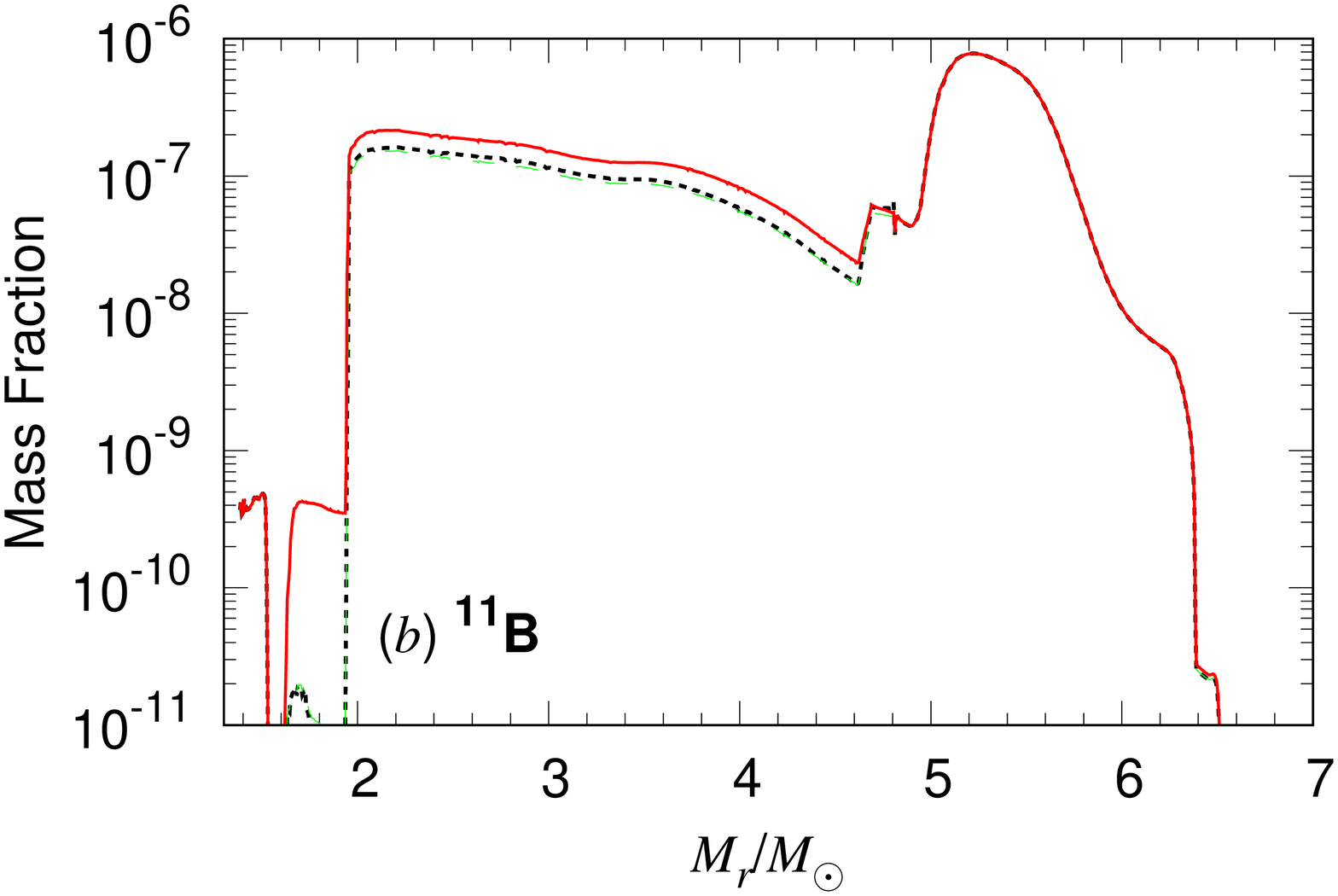}
\includegraphics[scale=0.30,angle=-90]{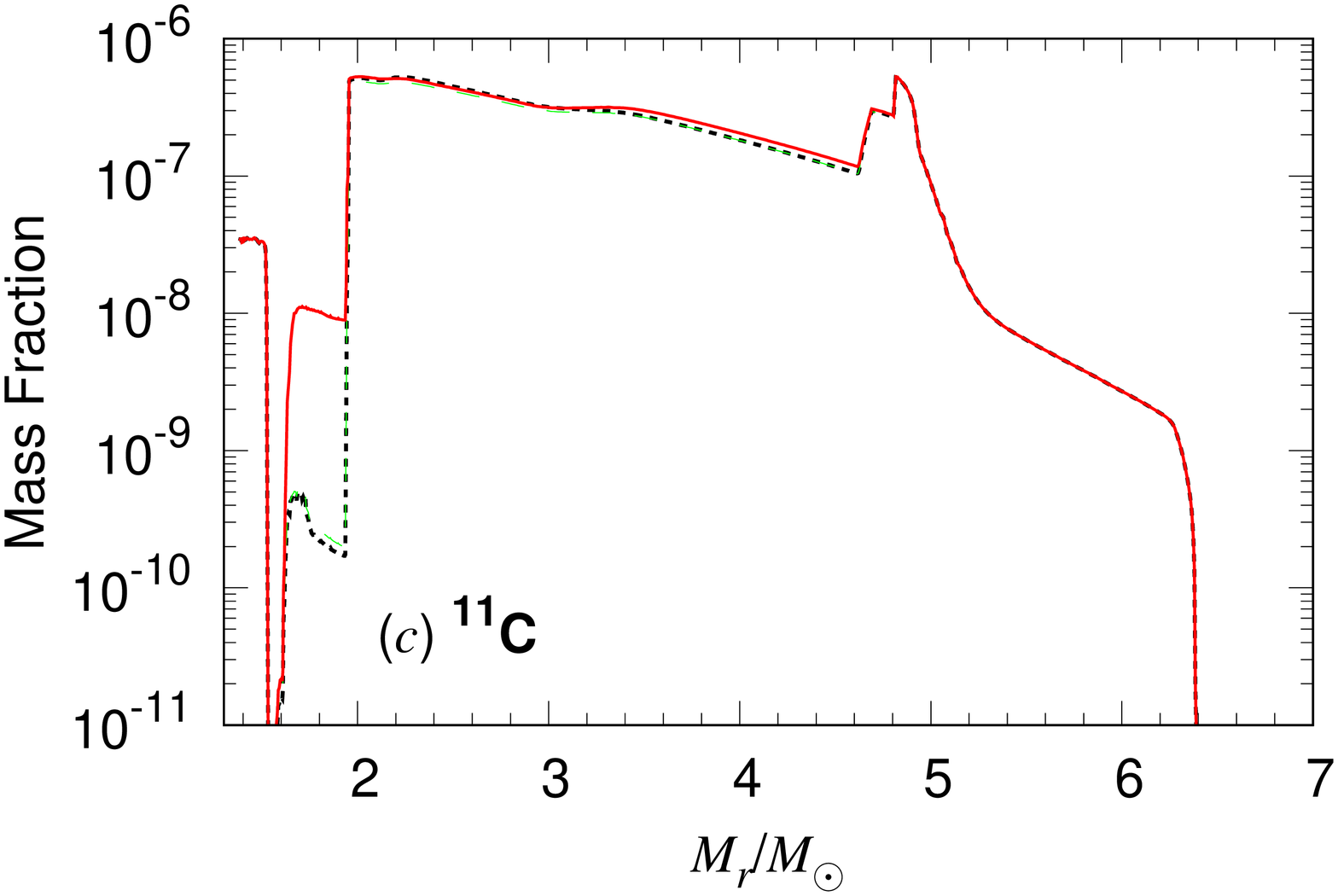}
\includegraphics[scale=0.30,angle=-90]{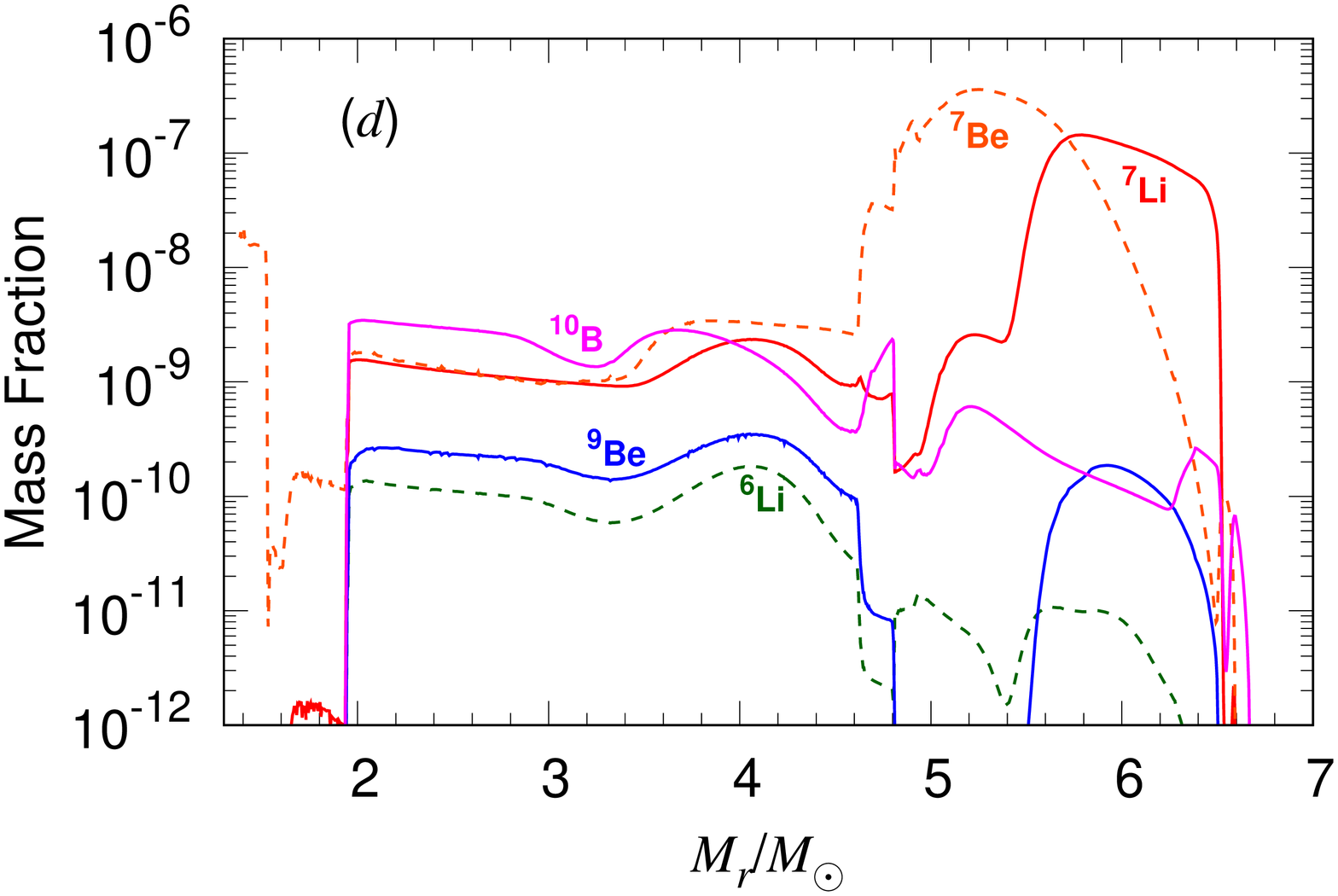}
\caption{(Color on line) The same as Fig. \ref{fig:fig12} for $M$ =20$M_{\odot}$.
\label{fig:fig13}}
\end{figure}

We calculate the nucleosynthesis of SN explosions of $M = $15 and 20 $M_\odot$ solar-metallicity stars.
The stellar evolution of these stars is calculated in the same manner of  \cite{YOU14}.
The 1D spherical SN explosions with the explosion energy of $1 \times 10^{51}$ erg is calculated for 100 s using a piecewise parabolic method code as in \cite{UN05}.
Then, we calculate the explosive nucleosynthesis during the SN explosions by post-processing with the neutrino process.
The adopted nuclei in the nuclear reaction network is as follows:
$^{1-3}$H, $^{3,4}$He, $^{6,7}$Li, $^{7,9}$Be, $^{8,10,11}$B, $^{11-16}$C, $^{13-18}$N, $^{14-20}$O, $^{17-22}$F, $^{18-24}$Ne. $^{21-26}$Na. $^{22-28}$Mg, $^{27-32}$Si, $^{27-34}$P, $^{30-37}$S, $^{32-38}$Cl, $^{34-43}$Ar, $^{36-45}$K, $^{38-48}$Ca, $^{40-49}$Sc, $^{42-51}$Ti, $^{44-53}$V, $^{46-55}$Cr, $^{48-57}$Mn, $^{50-61}$Fe, $^{51-62}$Co, $^{54-66}$Ni, $^{56-68}$Cu, $^{59-71}$Zn, $^{61-71}$Ga, $^{63-75}$Ge, $^{65-76}$As, $^{67-77}$Se, $^{70-79}$Br.
The total neutrino energy is set to be $3 \times 10^{53}$ erg.
The neutrino luminosity is assumed to decay exponentially in time with a time scale of 3 s and is equally partitioned among three flavors of neutrinos and antineutrinos.
The neutrino energy spectra are assumed to obey Fermi distributions with zero-chemical potentials.
The temperatures of $\nu_e$, $\bar{\nu}_e$ and $\nu_x$ =$\nu_{\mu,\tau}$ are set to be ($T_{\nu_e}$, $T_{\bar{\nu}_e}$, $T_{\nu_x}$) =(4, 4, 6) MeV.

In order to clarify the effect of the new cross section and branches of the $\nu$-$^{16}$O reactions, 
we consider three cases of the nucleosynthesis calculations.
In case 1, only single n, p, $\alpha$ and $\gamma$ emission channels are considered with previous HW92 cross sections \cite{Woos}.
New cross sections for $^{4}$He and $^{12}$C in Ref. \cite{YSK} are also used.
In case 2, the present new cross sections obtained with the SFO-tls are used, but with only single particle and $\gamma$ emission channels. 
In case 3, the multi-particle branches are included with the present new cross sections for $^{16}$O.

The production yields of $^{11}$B as well as $^{11}$C are estimated with the inclusion of the $^{16}$O ($\nu$, $\nu'\alpha$p) $^{11}$B and $^{16}$O ($\nu_e$, e$^{-}\alpha$p) $^{11}$C reactions for the SN explosion modes.
Calculated production yields are given in Table \ref{tab:table9}.
We see from Table \ref{tab:table9} that the production yields of the sum of $^{11}$B and $^{11}$C are enhanced by about 10$\%$ when the multi-particle channels are included.

Mass fraction distributions of $^{11}$B and $^{11}$C for the three cases as well as main elements are shown in Figs. 12 and 13 for the SN explosions of $M$ =15$M_{\odot}$ and $M$ = 20$M_{\odot}$ stars, respectively. 
$^{11}$B is produced mainly through $^{12}$C ($\nu$, $\nu$' p) $^{11}$B reaction in the O/Ne layer and $^{4}$He ($\nu$, $\nu$'p) $^{3}$H ($\alpha$, $\gamma$) $^{7}$Li ($\alpha$, $\gamma$) $^{11}$B in the He/C layers as we see from Figs. \ref{fig:fig12}(a), \ref{fig:fig12}(b) and Figs. \ref{fig:fig13}(a), \ref{fig:fig13}(b).
$^{11}$B is also produced in the O/Ne layer through $^{16}$O ($\nu$, $\nu$' $\alpha$p) $^{11}$B reaction, which enhances the mass fraction of $^{11}$B in the O/Ne layer as seen in Figs. \ref{fig:fig12}(b) and \ref{fig:fig13}(b). 
$^{11}$C is produced mainly through $^{12}$C ($\nu$, $\nu$'n) $^{11}$C reaction in the He/C and the O/Ne layers. 
It is produced also through $^{12}$C ($\nu_{e}$, e$^{-}$p) $^{11}$C reaction, and
$^{16}$O ($\nu_{e}$, e$^{-}$ $\alpha$p) $^{11}$C reaction in the O/Ne layer.
The cross section for the latter charged-current reaction on $^{16}$O is about 4$\%$ of that of the former one on $^{12}$C at $T$ = 4 MeV, while the mass fraction of $^{16}$O is larger than that of $^{12}$C by about 20 (8) times in the O/Ne layer for $M$ = 15 (20)$M_{\odot}$. 
This leads to similar production rates for the two charged-current reactions in the O/Ne layer in case of $M$ = 15$M_{\odot}$, 
but less production rate for the charged-current reaction on $^{16}$O compared with that on $^{12}$C in case of $M$ = 20$M_{\odot}$. 
Additional contributions come from a neutral current reaction $^{16}$O ($\nu$, $\nu$'$\alpha$n) $^{11}$C induced by $\nu_x$'s with higher temperature $T$ = 6 MeV.
These multi-particle emission channels lead to an enhancement of $^{11}$C in the O/Ne layer.    
As $^{11}$B is destroyed by $^{11}$B (p, $\alpha \alpha$) $^{4}$He reaction in the O/Ne layer, $^{11}$C is more produced than $^{11}$B in the O/Ne layer.     
Production yields of light elements, $^{6}$Li, $^{7}$Li, $^{7}$Be, $^{9}$Be and $^{10}$B, in the core-collapse SN explosions are also given in Table \ref{tab:table10}, 
which shows effects of the new cross sections and the multi-particle channels, and their mass fraction distributions in case 3 are shown in Figs. \ref{fig:fig12}(d) and \ref{fig:fig13}(d). 

\section{Summary}

We have studied $\nu$-induced reactions on $^{16}$O by shell-model calculations with the new Hamiltonian, SFO-tls \cite{SO}. 
Spin-dipole transitions are investigated as dominant contributions to the $\nu$-$^{16}$O reaction cross sections come from these transition strengths with the multipolarities, 0$^{-}$, 1$^{-}$ and 2$^{-}$.  
The spin-dipole strengths obtained with the SFO-tls are shifted toward lower energy region compared with the SFO \cite{SFO}.
The quenching of $g_A$ in the spin-dipole transitions is studied by muon capture reaction on $^{16}$O. Total muon capture rates obtained for the SFO and SFO-tls are found to agree with the experimental value within 10$\%$ with the use of the quenching factor f=0.95, which is used in the present calculations.    

Total as well as partial cross sections to various single- and multi-particle emission channels are evaluated by using the branching ratios obtained by Hauser-Feshbach statistical model. 
In the present calculation, the isospin conservation is fully taken into account.  

The total ($\nu$, e$^{-}$) cross section, partial cross sections for $\alpha$ emission channels, especially the $\alpha$p emission channel, in ($\nu$, e$^{-}$) and ($\nu$, $\nu$') reactions are found to be enhanced compared with the standard CRPA calculations. 
Effects of the inclusion of various multi-particle emission channels 
on nucleosynthesis of light elements in core-collapse SN explosions are investigated.
The $\alpha$p emission channels such as $^{16}O$ ($\nu$, $\nu$'$\alpha$p) $^{11}$B and $^{16}$O ($\nu$, e$^{-}$ $\alpha$p) $^{11}$C reactions are found to lead to an enhancement of the production yields of $^{11}$B and $^{11}$C 
by about 10$\%$ compared to the case with only single-nucleon emissions from $^{12}$C.
 
In Ref. \cite{Naka}, based on the present work, event spectra of the charged-current reactions as function of recoil energy of e$^{-}$/e$^{+}$ are discussed for future SN neutrino detection at the Super-Kamiokande.
Dependence on various SN neutrino spectra besides the Fermi distribution are also discussed.       
Thus, various applications to low energy neutrino detection and nucleosynthesis in stars can be carried out based on the present new reaction cross sections.  

\vspace*{0.7cm}
The authors would like to thank M. Sakuda and K. Nakazato  
for useful discussion on experiments and experimental plans at Super- and Hyper-Kamiokande and RCNP.
This work has been supported in part by JSPS KAKENHI Grant number JP15K05090.
K. T. was supported by the JSPS Overseas Research Fellowships. 




\newpage

\begin{table*}
\caption{\label{tab:table1}%
Total $\nu$-induced cross sections on $^{16}$O obtained with the SFO-tls, SFO and CRPA. The cross sections are given in units of 10$^{-42}$ cm$^{2}$ as function of the incoming neutrino energy $E_{\nu}$. Exponents are given in parentheses. 
}
\texttt{}
\begin{ruledtabular}
\begin{tabular}{c|ccc|ccc|ccc}
 &\multicolumn{3}{c|}{$^{16}$O($\nu$, e$^{-}$)$^{16}$F}&
\multicolumn{3}{c|}{$^{16}$O($\bar{\nu}$, e$^{+}$)$^{16}$N}&
\multicolumn{3}{c}{$^{16}$O($\nu$, $\nu$')$^{16}$O}\\
\hline
$E_{\nu}$ (MeV) & SFO-tls & SFO & CRPA & SFO-tls & SFO & CRPA & SFO-tls & SFO & CRPA\\
\hline
15.0 & 0.0 & 0.0 & 1.56(-6) & 1.33(-2) & 1.65(-2) & 2.53(-2) & 6.05(-4) & 6.06(-4) & 5.10(-4)\\ 
20.0 & 3.76(-2) & 2.51(-2) & 7.26(-3) & 1.38(-1) & 1.32(-1) & 1.81(-1) & 3.10(-2) & 2.61(-2) & 1.60(-2) \\ 
25.0 & 3.50(-1) & 2.48(-1) & 1.77(-1) & 7.09(-1) & 6.57(-1) & 8.90(-1) & 2.15(-1) & 1.79(-1) & 1.75(-1) \\
30.0 & 1.82(+0) & 1.38(+0) & 1.25(+0) & 2.56(+0) & 2.43(+0) & 2.94(+0) & 9.34(-1) & 8.05(-1) & 8.43(-1) \\
35.0 & 6.30(+0) & 5.16(+0) & 4.76(+0) & 6.61(+0) & 6.41(+0) & 7.26(+0) & 2.89(+0) & 2.61(+0) & 2.59(+0) \\
40.0 & 1.62(+1) & 1.40(+1) & 1.28(+1) & 1.38(+1) & 1.35(+1) & 1.48(+1) & 6.80(+0) & 6.34(+0) & 6.09(+0) \\
45.0 & 3.40(+1) & 3.05(+1) & 2.76(+1) & 2.48(+1) & 2.45(+1) & 2.64(+1) & 1.34(+1) & 1.27(+1) & 1.21(+1) \\
50.0 & 6.26(+1) & 5.74(+1) & 5.21(+1) & 4.02(+1) & 4.00(+1) & 4.29(+1) & 2.33(+1) & 2.24(+1) & 2.14(+1) \\
55.0 & 1.04(+2) & 9.72(+1) & 8.89(+1) & 6.03(+1) & 6.01(+1) & 6.46(+1) & 3.71(+1) & 3.60(+1) & 3.46(+1) \\
60.0 & 1.61(+2) & 1.52(+2) & 1.41(+2) & 8.51(+1) & 8.50(+1) & 9.17(+1) & 5.51(+1) & 5.37(+1) & 5.24(+1) \\
65.0 & 2.35(+2) & 2.24(+2) & 2.12(+2) & 1.14(+2) & 1.14(+2) & 1.25(+2) & 7.72(+1) & 7.57(+1) & 7.53(+1) \\
70.0 &  3.26(+2) & 2.93(+2) & 3.02(+2) & 1.47(+2) & 1.47(+2) & 1.63(+2) & 1.03(+2) & 1.02(+2) & 1.04(+2) \\
80.0 & 5.54(+2) & 5.37(+2) & 5.52(+2) & 2.20(+2) & 2.20(+2) & 2.57(+2) & 1.65(+2) & 1.64(+2) & 1.78(+2) \\
90.0 & 8.34(+2) & 8.15(+2) & 8.92(+2) & 2.97(+2) & 2.98(+2) & 3.77(+2) & 2.36(+2) & 2.35(+2) & 2.76(+2) \\
100.0 & 1.14(+3) & 1.13(+3) & 1.32(+3) & 3.74(+2) & 3.75(+2) & 5.18(+2) & 3.10(+2) & 3.09(+2) & 3.99(+2) \\
\end{tabular}
\end{ruledtabular}
\end{table*}

\begin{table*}
\caption{\label{tab:table2}%
Averaged total $\nu$-induced cross sections on $^{16}$O over Fermi distributions of neutrino spectra with temperatures $T$. The cross sections are given in units of 10$^{-42}$ cm$^{2}$ as function of the temperature. Exponents are given in parentheses. 
}
\texttt{}
\begin{ruledtabular}
\begin{tabular}{c|ccc|ccc|ccc}
 &\multicolumn{3}{c|}{$^{16}$O($\nu$, e$^{-}$)$^{16}$F}&
\multicolumn{3}{c|}{$^{16}$O($\bar{\nu}$, e$^{+}$)$^{16}$N}&
\multicolumn{3}{c}{$^{16}$O($\nu$, $\nu$')$^{16}$O}\\
\hline
$T$ (MeV) & SFO-tls & SFO & CRPA & SFO-tls & SFO & CRPA & SFO-tls & SFO & CRPA\\
\hline
2 & 7.83(-3) & 5.60(-4) &  & 2.32(-3) & 2.38(-3) &  & 5.30(-4) & 4.52(-4) & \\
3 & 2.97(-2) & 2.35(-2) & & 4.42(-2) & 4.27(-2) & & 1.52(-2) & 1.42(-2) &  \\
4 & 2.59(-1) & 2.19(-1) & 1.91(-1) & 2.70(-1) & 2.62(-1) & & 1.15(-1) & 1.11(-1) & \\
5 & 1.13(+0) & 9.96(-1) & & 9.44(-1) & 9.22(-1) & 1.05(+0) & 4.61(-1) & 4.53(-1) & \\
6 & 3.36(+0) & 3.03(+0) & & 2.38(+0) & 2.34(+0) & 1.28(+0) & 1.28(+0) & \\
7 & 7.79(+0) & 7.17(+0) & & 4.88(+0) & 4.82(+0) & 2.81(+0) & 2.84(+0) & \\
8 & 1.53(+1) & 1.43(+1) & 1.37(+1) & 8.66(+0) & 8.59(+0) & 9.63(+0) & 5.29(+0) & 5.38(+0) & 5.19(+0) \\
9 & 2.66(+1) & 2.51(+1) & & 1.39(+1) & 1.38(+1) & & 8.86(+0) & 9.08(+0) &  \\
10 & 4.23(+1) & 4.02(+1) & & 2.06(+1) & 2.05(+1) & & 1.37(+1) & 1.41(+1) & \\ 
\end{tabular}
\end{ruledtabular}
\end{table*}

\begin{table*}
\caption{\label{tab:table3}%
Partial cross sections of $^{16}$O ($\nu$, e$^{-}$ X) for various channels obtained with the SFO-tls. The cross sections are given in units of 10$^{-42}$ cm$^{2}$ as function of the incoming neutrino energy $E_{\nu}$. Exponents are given in parentheses. 
}
\texttt{}
\begin{ruledtabular}
\begin{tabular}{c|ccccccc}
$E_{\nu}$ (MeV) & p & d, pn & pp & $^{3}$He & $\alpha$ & $^{3}$He p & $\alpha$p \\
\hline
15.0 & 0.00 & 0.0 & 0.0 & 0.0 & 0.0 & 0.0 & 0.0 \\ 
20.0 & 3.76(-2) & 0.0 & 0.0 & 0.0 & 0.0 & 0.0 & 0.0 \\ 
25.0 & 3.50(-1) & 1.94(-6) & 5.30(-4) & 2.40(-6) & 8.82(-6) & 2.14(-9) & 7.67(-8) \\
30.0 & 1.78(+0) & 3.78(-3) & 1.93(-2) & 4.41(-3) & 4.96(-3) & 9.23(-4) & 4.07(-3) \\
35.0 & 5.72(+0) & 5.32(-2) & 2.42(-1) & 6.47(-2) & 5.96(-2) & 3.76(-2) & 1.16(-1) \\
40.0 & 1.40(+1) & 1.85(-1) & 9.05(-1) & 2.33(-1) & 1.98(-1) & 1.75(-1) & 5.09(-1) \\
45.0 & 2.84(+1) & 4.34(-1) & 2.26(+0) & 5.63(-1) & 4.54(-1) & 4.87(-1) & 1.37(+0) \\
50.0 & 5.12(+1) & 8.38(-1) & 4.59(+0) & 1.12(+0) & 8.65(-1) & 1.06(+0) & 2.91(+0) \\
55.0 & 8.39(+1) & 1.44(+0) & 8.18(+0) & 1.96(+0) & 1.47(+0) & 1.99(+0) & 5.35(+0) \\
60.0 & 1.28(+2) & 2.27(+0) & 1.33(+1) & 3.16(+0) & 2.32(+0) & 3.34(+0) & 8.88(+0) \\
65.0 & 1.84(+2) & 3.36(+0) & 2.02(+1) & 4.74(+0) & 3.42(+0) & 5.18(+0) & 1.36(+1) \\
70.0 & 2.52(+2) & 4.72(+0) & 2.89(+1) & 6.75(+0) & 4.82(+0) & 7.52(+0) & 1.97(+1) \\
80.0 & 4.22(+2) & 8.26(+0) & 5.22(+1) & 1.21(+1) & 8.50(+0) & 1.37(+1) & 3.60(+1) \\
90.0 & 6.26(+2) & 1.27(+1) & 8.26(+1) & 1.89(+1) & 1.33(+1) & 2.16(+1) & 5.70(+1) \\
100.0 & 8.48(+2) & 1.78(+1) & 1.18(+2) & 2.70(+1) & 1.89(+1) & 3.04(+1) & 8.10(+1) \\
\end{tabular}
\end{ruledtabular}
\end{table*}

\begin{table*}
\caption{\label{tab:table4}%
Partial cross sections of $^{16}$O ($\bar{\nu}$, e$^{+}$ X) for various channels obtained with the SFO-tls. The cross sections are given in units of 10$^{-42}$ cm$^{2}$ as function of the incoming neutrino energy $E_{\nu}$. Exponents are given in parentheses. 
}
\texttt{}
\begin{ruledtabular}
\begin{tabular}{c|cccccc}
$E_{\nu}$ (MeV) & $^{16}$O ($\bar{\nu}$, e$^{+}$) $^{16}$N$_{g.s.}$ & n & d, pn & $^{3}$H & $\alpha$ & $\alpha$n \\
\hline
15.0 & 1.33(-2) & 2.93(-5) & 0.0 & 0.0 & 0.0 & 0.0  \\ 
20.0 & 1.26(-1) & 1.27(-2) & 8.89(-7) & 0.0 & 4.19(-7) & 0.0 \\ 
25.0 & 4.74(-1) & 2.31(-1) & 1.46(-3) & 2.39(-4) & 1.69(-3) & 3.95(-5) \\
30.0 & 1.22(+0) & 1.24(+0) & 3.76(-2) & 9.78(-3) & 4.61(-2) & 3.32(-3) \\
35.0 & 2.54(+0) & 3.67(+0) & 1.41(-1) & 4.67(-2) & 1.87(-1) & 2.12(-2) \\
40.0 & 4.55(+0) & 8.19(+0) & 3.31(-1) & 1.30(-1) & 4.73(-1) & 6.11(-2) \\
45.0 & 7.35(+0) & 1.54(+1) & 6.27(-1) & 2.85(-1) & 9.60(-1) & 1.32(-1) \\
50.0 & 1.10(+1) & 2.56(+1) & 1.04(+0) & 5.29(-1) & 1.69(+0) & 2.37(-1) \\
55.0 & 1.53(+1) & 3.92(+1) & 1.59(+0) & 8.77(-1) & 2.71(+0) & 3.77(-1) \\
60.0 & 2.04(+1) & 5.62(+1) & 2.27(+0) & 1.33(+0) & 4.02(+0) & 5.50(-1) \\
65.0 & 2.60(+1) & 7.62(+1) & 3.08(+0) & 1.89(+0) & 5.63(+0) & 7.55(-1) \\
70.0 & 3.20(+1) & 9.90(+1) & 4.00(+0) & 2.53(+0) & 7.51(+0) & 9.89(-1) \\
80.0 & 4.45(+1) & 1.50(+2) & 6.10(+0) & 4.04(+0) & 1.19(+1) & 1.53(+0) \\
90.0 & 5.66(+1) & 2.06(+2) & 8.39(+0) & 5.69(+0) & 1.68(+1) & 2.15(+0) \\
100.0 & 6.75(+1) & 2.62(+2) & 1.07(+1) & 7.32(+0) & 2.16(+1) & 2.82(+0) \\
\end{tabular}
\end{ruledtabular}
\end{table*}

\begin{table*}
\caption{\label{tab:table5}%
The same as in Table III and IV for partial cross sections of $^{16}$O ($\nu$, $\nu$' X) for various channels obtained with the SFO-tls. 
}
\texttt{}
\begin{ruledtabular}
\begin{tabular}{c|cccccccccc}
$E_{\nu}$ (MeV) & $\gamma$ & n & p & d, pn & pp & $^{3}$H & $^{3}$He\\
 & $\alpha$ & $\alpha$n & $\alpha$p \\
\hline
15.0 & 1.35(-7) & 0.0 & 6.05(-4) & 0.0 & 0.0 & 0.0 & 0.0 \\
& 1.35(-25) & 0.0 & 0.0\\ 
20.0 & 1.02(-5) & 1.93(-4) & 3.08(-2) & 0.0 & 0.0 & 0.0 & 0.0 \\
&1.47(-10) & 0.0 & 0.0\\ 
25.0 & 9.97(-5) & 1.63(-2) & 1.98(-1) & 2.88(-5) & 1.09(-5) & 2.46(-9) & 1.82(-5)\\
 & 1.74(-6) & 0.0 & 4.30(-8) \\
30.0 & 5.54(-4) & 1.29(-1) & 7.76(-1) & 7.92(-3) & 5.71(-3) & 1.82(-3) & 4.19(-3)\\
 & 4.14(-3) & 3.76(-5) & 4.83(-3) \\
35.0 & 1.83(-3) & 4.80(-1) & 2.16(+0) & 6.43(-2) & 4.74(-2) & 1.48(-2) & 3.56(-2) \\
& 3.17(-2) & 6.86(-4) & 5.39(-2) \\
40.0 & 4.46(-3) & 1.22(+0) & 4.76(+0) & 2.06(-1) & 1.51(-1) & 4.52(-2) & 1.18(-1) \\
& 9.55(-2) & 3.74(-3) &1.99(-1) \\
45.0 & 9.00(-3) & 2.49(+0) & 9.01(+0) & 4.68(-1) & 3.41(-1) & 9.96(-2) & 2.75(-1) \\
& 2.07(-1) & 1.09(-2) & 4.96(-1) \\
50.0 & 1.60(-2) & 4.44(+0) & 1.52(+1) & 8.87(-1) & 6.44(-1) & 1.85(-1) & 5.36(-1) \\
& 3.81(-1) & 2.43(-2) & 9.97(-1) \\
55.0 & 2.60(-2) & 7.17(+0) & 2.37(+1) & 1.49(+0) & 1.08(+0) & 3.06(-1) & 9.23(-1) \\
& 6.27(-1) & 4.53(-2) & 1.75(+0) \\
60.0 & 3.94(-2) & 1.07(+1) & 3.45(+1) & 2.31(+0) & 1.68(+0) & 4.69(-1) & 1.45(+0) \\
& 9.57(-1) & 7.48(-2) & 2.80(+0) \\
65.0 & 5.66(-2) & 1.52(+1) & 4.77(+1) & 3.36(+0) & 2.43(+0) & 6.75(-1) & 2.14(+0)\\
 & 1.37(+0) & 1.13(-1) & 4.17(+0) \\
70.0 & 7.78(-2) & 2.04(+1) & 6.29(+1) & 6.64(+0) & 3.35(+0) & 9.24(-1) & 2.97(+0)\\
 & 1.87(+0) & 1.60(-1) & 5.85(+0) \\
80.0 & 1.34(-1) & 3.28(+1) & 9.86(+1) & 7.82(+0) & 5.63(+0) & 1.54(+0) & 5.06(+0) \\
& 3.11(+0) & 2.78(-1)& 1.01(+1) \\
90.0 & 2.10(-1) & 4.68(+1) & 1.38(+2) & 1.17(+1) & 8.36(+0) & 2.27(+0) & 7.57(+0) \\
& 4.58(+0) & 4.23(-1) & 1.53(+1) \\
100.0 & 3.12(-1) & 6.11(+1) & 1.79(+2) & 1.59(+1) & 1.13(+1) & 3.06(+0) & 1.03(+1)\\
 & 6.20(+0) & 5.85(-1) & 2.11(+1) \\
\end{tabular}
\end{ruledtabular}
\end{table*}

\begin{table*}
\caption{\label{tab:table6}%
Averaged cross sections of $^{16}$O($\nu$, e$^{-}$ X) folded over Fermi distributions of neutrino spectra with temperatures $T$ = 4 MeV and 8 MeV obtained with the SFO-tls and CRPA. The cross sections are given in units of 10$^{-42}$ cm$^{2}$ as function of the temperature. Exponents are given in parentheses. 
In case of X=p, values in the parentheses are those for the transitions to $^{15}$O$_{g.s.}$.
}
\texttt{}
\begin{ruledtabular}
\begin{tabular}{c|cc|cc}
 Neutrino reactions &\multicolumn{2}{c|}{$T$ =4 MeV}&
\multicolumn{2}{c}{$T$ = 8 MeV}\\
\hline
 & SFO-tls & CRPA & SFO-tls & CRPA \\
\hline
total & 2.59(-1) & 1.91(-1) & 1.53(+1) & 1.37(+1) \\
$^{16}$O($\nu$, e$^{-}$ p )$^{15}$O & 2.29(-1) & 1.62(-1) (1.21(-1))& 1.22(+1) & 9.56(+0) (6.37(+0) \\
$^{16}$O($\nu$, e$^{-}$ np)$^{14}$O & 2.36(-3) & 3.92(-4) & 2.08(-1) & 1.76(-1) \\
$^{16}$O($\nu$, e$^{-}$ pp)$^{14}$N & 1.21(-2) & 2.61(-2) & 1.22(+0) & 3.26(+0) \\
$^{16}$O($\nu$, e$^{-}$ $^{3}$He)$^{13}$N & 3.04(-3) &   & 2.90(-1) &   \\
$^{16}$O($\nu$, e$^{-}$ $\alpha$)$^{12}$N & 2.52(-3)&1.16(-3)& 2.16(-1)& 1.31(-1)\\
$^{16}$O($\nu$, e$^{-}$ $^{3}$He p)$^{12}$C & 2.47(-3) &   & 3.00(-1) &  \\
$^{16}$O($\nu$, e$^{-}$ $\alpha$ p)$^{11}$C & 7.00(-3) & 2.17(-3) & 8.06(-1) & 5.66(-1) \\
\end{tabular}
\end{ruledtabular}
\end{table*}

\begin{table*}
\caption{\label{tab:table7}%
Averaged cross sections of $^{16}$O($\bar{\nu}$, e$^{+}$ X) folded over Fermi distributions of neutrino spectra with temperatures $T$ = 5 MeV and 8 MeV obtained with the SFO-tls and CRPA. The cross sections are given in units of 10$^{-42}$ cm$^{2}$ as function of the temperature. Exponents are given in parentheses. 
In case of X=n, values in the parentheses are those for the transitions to $^{15}$N$_{g.s.}$.
}
\texttt{}
\begin{ruledtabular}
\begin{tabular}{c|cc|cc}
 Neutrino reactions &\multicolumn{2}{c|}{$T$ =5 MeV}&
\multicolumn{2}{c}{$T$ = 8 MeV}\\
\hline
 & SFO-tls & CRPA & SFO-tls & CRPA \\
\hline
total & 9.44(-1) & 1.05(+0) & 8.66(+0) & 9.63(+0) \\
$^{16}$O($\bar{\nu}$, e$^{+}$)$^{16}$N & 3.51(-1)& 4.94(-1) (3.47(-1))& 2.37(+0)& 4.05(+0) (2.15(+0)) \\
$^{16}$O($\bar{\nu}$, e$^{+}$ n)$^{15}$N & 5.29(-1)& 6.71(-1) (5.24(-1))& 5.49(+0) & 6.71(+0) (4.81(+0))\\
$^{16}$O($\bar{\nu}$, e$^{+}$ np)$^{14}$C & 1.99(-2) & 4.56(-3) & 2.20(-1) & 1.38(-1) \\
$^{16}$O($\bar{\nu}$, e$^{+}$ nn)$^{14}$N & 1.09(-3) & 5.50(-3) & 1.42(-2) & 1.81(-1) \\
$^{16}$O($\bar{\nu}$, e$^{+}$ $^{3}$H)$^{13}$C & 8.64(-3) &   & 1.19(-1) &   \\
$^{16}$O($\bar{\nu}$, e$^{+}$ $\alpha$)$^{12}$B & 2.99(-2)&1.07(-2)& 3.75(-1)& 1.91(-1)\\
$^{16}$O($\bar{\nu}$, e$^{+}$ $\alpha$ n)$^{11}$B & 3.78(-3) & 6.20(-3) & 4.97(-2) & 2.16(-1) \\
\end{tabular}
\end{ruledtabular}
\end{table*}

\begin{table*}
\caption{\label{tab:table8}%
Averaged cross sections of $^{16}$O($\nu$, $\nu$'X) folded over Fermi distributions of neutrino spectra with temperatures $T$ =4 MeV, 6 MeV and 8 MeV obtained with the SFO-tls and CRPA. The cross sections are given in units of 10$^{-42}$ cm$^{2}$ as function of the temperature. Exponents are given in parentheses. 
In case of X=n (X=p), values in the parentheses are those for the transitions to $^{15}$O$_{g.s.}$  ($^{15}$N$_{g.s.}$).
}
\texttt{}
\begin{ruledtabular}
\begin{tabular}{c|c|c|cc}
 Neutrino reactions & \multicolumn{1}{c|}{$T$ = 4 MeV} & \multicolumn{1}{c|}{$T$ = 6 MeV} & \multicolumn{2}{c}{$T$ = 8 MeV}\\
\hline
 & SFO-tls & SFO-tls & SFO-tls & CRPA  \\
\hline
total & 1.15(-1) & 1.28(+0) & 5.29(+0) & 5.19(+1) \\
$^{16}$O($\nu$, $\nu$'$\gamma$)$^{16}$O & 7.16(-5) & 8.73(-4) & 3.88(-3) & 3.19(-3) \\
$^{16}$O($\nu$, $\nu$'n)$^{15}$O  &  1.81(-2) & 2.31(-1) & 1.00(+0) & 1.32(+0) (9.73(-1))\\
$^{16}$O($\nu$, $\nu$'p)$^{15}$N & 8.66(-2) & 8.66(-1) & 3.40(+0) & 3.14(+0) (1.85(+0))\\
$^{16}$O($\nu$, $\nu$'np)$^{14}$N &  2.66(-3) & 4.37(-2) & 2.10(-1) & 4.40(-1) \\
$^{16}$O($\nu$, $\nu$'pp)$^{14}$C & 1.94(-3) & 3.17(-2) & 1.52(-1) & 8.35(-2) \\ 
$^{16}$O($\nu$, $\nu$'$^{3}$H)$^{13}$N & 5.77(-4) & 9.12(-3) & 7.45(-2) &     \\
$^{16}$O($\nu$, $\nu$'$^{3}$He)$^{13}$C & 1.55(-3) & 2.65(-2) & 1.31(-1) &  \\
$^{16}$O($\nu$, $\nu$'$\alpha$)$^{12}$C & 1.22(-3) & 1.89(-2) & 8.77(-2)&   \\
$^{16}$O($\nu$, $\nu$'$\alpha$ n)$^{11}$C & 5.38(-5) & 1.17(-3) & 6.43(-3)  & 3.88(-2)  \\
$^{16}$O($\nu$, $\nu$'$\alpha$ p)$^{11}$B & 2.65(-3) & 4.89(-2) & 2.50(-1) & 9.15(-2) \\
\end{tabular}
\end{ruledtabular}
\end{table*}

\begin{table*}
\caption{\label{tab:table9}%
Production yields of $^{11}$B and $^{11}$C in SN explosions for progenitor mass of $M$ = 15$M_{\odot}$ and 20$M_{\odot}$. 
}
\texttt{}
\begin{ruledtabular}
\begin{tabular}{c|ccc|ccc}
 &\multicolumn{3}{c|}{$M$=15$M_{\odot}$}&
\multicolumn{3}{c}{$M$=20$M_{\odot}$}\\
\hline
Production yield (10$^{-7}M_{\odot}$) & case 1 & case 2 & case 3 & case 1 & case 2 & case 3 \\
\hline
M($^{11}$B)  & 2.94 & 2.92 &  3.13 & 6.77 &  6.58 & 7.66 \\
M($^{11}$C)  & 2.80 & 2.71 & 3.20 & 9.33 & 8.91 & 9.64 \\
M($^{11}$B +$^{11}$C)  & 5.74 & 5.62 & 6.33 & 16.10 & 15.49 & 17.29 \\ 
\end{tabular}
\end{ruledtabular}
\end{table*}

\begin{table*}
\caption{\label{tab:table10}%
Production yields of $^{6}$Li, $^{7}$Li, $^{7}$Be, $^{9}$Be and $^{10}$B in SN explosions for progenitor mass of $M$ = 15$M_{\odot}$ and 20$M_{\odot}$. 
They are given in units of $M_{\odot}$. Exponents are given in parentheses.
}
\texttt{}
\begin{ruledtabular}
\begin{tabular}{c|ccc|ccc}
 &\multicolumn{3}{c|}{$M$=15$M_{\odot}$}&
\multicolumn{3}{c}{$M$=20$M_{\odot}$}\\
\hline
Production yield ($M_{\odot}$) & case 1 & case 2 & case 3 & case 1 & case 2 & case 3 \\
\hline
M($^{6}$Li)  & 4.29(-11) & 4.16(-11) & 4.39(-11) & 2.33(-10) & 2.19(-10) & 2.90(-10) \\
M($^{7}$Li)  & 7.23(-8) & 7.23(-8) & 7.09(-8) & 1.11(-7) & 1.10(-7) & 1.02(-7) \\
M($^{7}$Be)  & 2.33(-7) & 2.33(-7) & 2.33(-7) & 2.57(-7) & 2.57(-7) & 2.61(-7) \\ 
M($^{9}$Be)  & 9.62(-11) & 9.50(-11) & 1.01(-10) & 4.12(-10) & 3.81(-10) & 6.93(-10) \\
M($^{10}$B) & 2.18(-9) & 2.12(-9) & 2.42(-9) & 6.71(-9) & 6.15 (-9) & 6.47(-9) \\
\end{tabular}
\end{ruledtabular}
\end{table*}

\end{document}